\documentclass[aps,10pt,pra,superscriptaddress,twocolumn,nofootinbib]{revtex4-2}
\usepackage[utf8]{inputenc}
\usepackage{CJKutf8}
\usepackage{multirow}
\usepackage{amsmath,bm}
\usepackage{times}
\usepackage{float}
\usepackage{listings}
\usepackage{mdframed}
\usepackage[dvipsnames]{xcolor}

\usepackage{tikz}

\usepackage{amssymb}
\usepackage{amsthm}

\newtheorem{theorem}{Theorem}

\newtheorem{hypothesis}[theorem]{Null Hypothesis}

\DeclareMathOperator*{\argmin}{\text{argmin}}

\usepackage{braket}
\usepackage{graphicx}
\usepackage{subcaption}
\usepackage{bbold}
\usepackage[hyperindex,breaklinks,colorlinks=true,citecolor=blue,urlcolor=blue]{hyperref}
\usepackage{hhline}

\newcommand{\F}{\mathcal{F}}
\renewcommand{\H}{\mathcal{H}}

\newcommand{\Q}{\mathcal{Q}}
\renewcommand{\P}{\mathcal{P}}
\newcommand{\Pf}{\mathfrak{P}}
\renewcommand{\S}{\mathcal{S}}

\newcommand{\bfv}{\mathbf{v}}
\newcommand{\bfN}{\bm{N}}
\newcommand{\bmF}{\bm{F}}

\newcommand{\vecB}{\vec{\beta}}
\newcommand{\vecf}{\vec{f}}
\newcommand{\vecP}{\vec{P}}
\newcommand{\PCHSH}{\vecP_\text{CHSH}}
\newcommand{\PCGLMP}{\vecP_\text{CGLMP}}

\newcommand{\Neg}{\mathcal{N}}

\newcommand{\Ntot}{N_\text{tot}}
\newcommand{\Nint}{N_\text{blk}}
\newcommand{\Nest}{N_\text{est}}
\newcommand{\Ntest}{N_\text{test}}

\newcommand{\Tp}{\mbox{\tiny T}}
\newcommand{\prot}{^{\text{(prot)}}}
\newcommand{\mart}{^{\text{(mart)}}}
\newcommand{\pbr}{^{\text{(pbr)}}}
\newcommand{\reg}{_{\text{reg}}}
\newcommand{\rhoS}{\rho_\text{\tiny SWAP}}
\newcommand{\iid}{{\em i.i.d}}

\definecolor{nred}{rgb}{0.9,0.1,0.1}
\definecolor{nblack}{rgb}{0,0,0}
\definecolor{nblue}{rgb}{0.2,0.2,0.8}
\definecolor{ngreen}{rgb}{0.2,0.6,0.2}
\definecolor{carnelian}{rgb}{0.7, 0.11, 0.11}

\newcommand{\proj}[1]{\ket{#1}\!\bra{#1}}

\newcommand{\ExpVal}[1]{\left\langle #1\right\rangle}
\newcommand{\tr}{{\rm tr}}
\newcommand{\ICHSH}{I_\text{\tiny CHSH}}
\newcommand{\SCHSH}{\S_\text{\tiny CHSH}}
\newcommand{\ICGLMP}{I_\text{\tiny CGLMP3}}
\newcommand{\SCGLMP}{\S_\text{\tiny CGLMP3}}
\newcommand{\SMermin}{\S_\text{\tiny Mermin}}

\newcommand{\Id}{\mathbb{1}}

\usepackage[capitalize]{cleveref}

\renewcommand{\L}{\mathcal{L}}

\begin{document}

\begin{CJK*}{UTF8}{bsmi}

\title{Device-independent certification of desirable properties with a confidence interval}

\author{Wan-Guan Chang}
\affiliation{Department of Physics and Center for Quantum Frontiers of Research \& Technology (QFort), National Cheng Kung University, Tainan 701, Taiwan}
\affiliation{Institute of Information Science, Academia Sinica, Taiwan}
\affiliation{Physics Division, National Center for Theoretical Sciences, Taipei 10617, Taiwan}

\author{Kai-Chun Chen}
\affiliation{Department of Physics and Center for Quantum Frontiers of Research \& Technology (QFort), National Cheng Kung University, Tainan 701, Taiwan}

\author{Kai-Siang Chen}
\affiliation{Department of Physics and Center for Quantum Frontiers of Research \& Technology (QFort), National Cheng Kung University, Tainan 701, Taiwan}

\author{Shin-Liang Chen}
\affiliation{Department of Physics, National Chung Hsing University, Taichung 40227, Taiwan}
\affiliation{Physics Division, National Center for Theoretical Sciences, Taipei 10617, Taiwan}

\author{Yeong-Cherng Liang}
\email{ycliang@mail.ncku.edu.tw}
\affiliation{Department of Physics and Center for Quantum Frontiers of Research \& Technology (QFort), National Cheng Kung University, Tainan 701, Taiwan}
\affiliation{Physics Division, National Center for Theoretical Sciences, Taipei 10617, Taiwan}

\begin{abstract}
In the development of quantum technologies, a reliable means for characterizing quantum devices, be it a measurement device, a state-preparation device, or a transformation device, is crucial. However, the conventional approach based on, for example, quantum state tomography or process tomography relies on assumptions that are often not necessarily justifiable in a realistic experimental setting. While the device-independent approach to this problem bypasses the shortcomings above by making only minimal, justifiable assumptions, most of the theoretical proposals to date only work in the idealized setting where independent and identically distributed (i.i.d.) trials are assumed. Here, we provide a versatile solution for rigorous device-independent certification that does not rely on the i.i.d. assumption. Specifically, we describe how the prediction-based-ratio (PBR) protocol and martingale-based protocol developed for hypothesis testing can be applied in the present context to achieve a device-independent certification of desirable properties with confidence interval. To illustrate the versatility of these methods, we demonstrate how we can use them to certify---with finite data---the underlying negativity, Hilbert space dimension, entanglement depth, and fidelity to some target pure state. In particular, we give examples showing how the amount of certifiable negativity and fidelity scales with the number of trials, and how many experimental trials one needs to certify a qutrit state space, or the presence of genuine tripartite entanglement. Overall, we have found that the PBR protocol and the martingale-based protocol often offer similar performance, even though the former does have to presuppose any witness (Bell-like inequality). In contrast, our findings also show that the performance of the martingale-based protocol may be severely affected by one's choice of the Bell-like inequality. Intriguingly, a Bell function useful for self-testing does not necessarily give the optimal confidence-gain rate for certifying the fidelity to the corresponding target state. 
\end{abstract}
\date{\today}
\maketitle

\section{Introduction}

The proper analysis of quantum experiments is an indispensable part in the development of quantum technologies. However, it is not trivial to reliably characterize a quantum setup, which may include, e.g., measurement devices and state-preparation devices. Moreover, imperfections in the experimental setup can easily result in a mismatch~\cite{{RFB+12,MKS+13,van_Enk2013_imperfect_experiment}} between the characterization tools developed for an idealized situation and an actual experimental situation. However, we can circumvent this problem by the so-called ``device-independent approach"~\cite{Scarani12,BCP+14}. In quantum information, the term ``device-independent" (DI) was first coined~\cite{DI2006} in the task of quantum key distribution~\cite{BB84,Ekert91,GGTZ02}, even though the idea was already conceived independently, but implicitly in~\cite{MY98,MY04}. 

In a nutshell, the DI approach is a framework for analyzing physical systems without relying on any assumption about the degrees of freedom measured. Its basis is Bell-nonlocality~\cite{Bell64,BCP+14}, which shows that no local-hidden-variable theory (LHV) can reproduce {\em all} quantum predictions, even though {\em no} further assumption is made about the details of such a theory. For example, it is known that the violation of Bell inequalities~\cite{Bell64} obtained by locally measuring a shared state implies~\cite{Werner:PRA:1989} shared entanglement~\cite{HHHH09}, which is a powerful resource in many quantum information processing tasks. More generally, many other desirable properties of the underlying state~\cite{BPA+08,BGLP11,MBL+13,TMG15,LRB+15,AB19,CKZ+21,CBLC16, CBLC18}, measurements~\cite{CBLC16, CBLC18,Bancal:PRL:2018,Renou:PRL:2018,QBW+19,WBSS20,CMBC21}, and channel~\cite{CKZ+21,Sekatski2018,SNI+23} may be derived directly from the observation of a Bell-inequality-violating correlation between measurement outcomes. Recently, the DI approach has been also been incorporated into the security analysis of quantum secure direct communication, see, e.g.,~\cite{ZSL20} and references therein.

However, due to statistical fluctuations, even when the experimental trials are independent and identically distributed (\iid), relative frequencies of the measurement outcomes obtained from a Bell experiment do not faithfully represent the underlying distribution. In particular, such raw distributions estimated from the experimental results typically~\cite{Bernhard2014,DIestimation2014,SBSL16} lead to a violation of the nonsignaling conditions \cite{PR94,BLM+05}, which is a prerequisite for the analysis in~\cite{BGLP11,MBL+13,TMG15,LRB+15,CKZ+21,CBLC16, CBLC18,Bancal:PRL:2018,Renou:PRL:2018,AB19,QBW+19,WBSS20,Sekatski2018,CMBC21,SNI+23}. In other words, statistical fluctuations render the many theoretical tools developed for such a purpose inapplicable. To address this issue, some {\em ad hoc} methods~\cite{Bernhard2014,DIestimation2014,SBSL16} have been proposed to regularize the relative frequencies obtained to ensure that the resulting distribution satisfy the nonsignaling conditions. In~\cite{{LRZ+18}}, a more in-depth discussion was provided and two better-motivated regularization methods were proposed.

While these more recent attempts do provide a point estimator that fits within the framework of the usual DI analysis, they are still problematic in two aspects. Firstly, they do not provide any confidence region associated with the estimate.  However, any real experiment necessarily  involves only a finite number of experimental trials. Therefore a useful analysis should provide not only an estimate but also an indication of the reliability of such an estimate. In many of the Bell experiments reported \cite{{ADG82,TBZ+98,WJS+98,Rowe2001}}, this is achieved by reporting the standard deviations of Bell violations. However, for finite, especially a relatively small number of trials, the central limit theorem is not warranted, so the usual interpretation of standard deviations may become dubious. Secondly, these usual approaches and those that provide a DI point estimator~\cite{Bernhard2014,DIestimation2014,SBSL16,LRZ+18} implicitly assumes that the experimental trials are \iid, and hence free of the memory effect \cite{{memoryeffect,Gill2003}} (see more discussions in \cite{BCP+14,Hensen15,Shalm2015,GVW+15}). Again, in a realistic experimental setting, the {\iid} assumption may be difficult to justify.

For the tasks of DI randomness expansion \cite{Colbeck2006,PMG+10} and DI quantum key distribution \cite{ABG+07,PAB+09}, specific tools~\cite{DIrandom2013,DIrandom2018,Bierhorst:2018aa,Bourdoncle_2019,ZFK20,DIrandom2020,Metger22,DIQKD2019,DIQKD2020} have been developed to overcome the above problems. Here, we are interested in providing a general solution to other device-independent certification tasks\footnote{Note that the same task is called device-independent verification in~\cite{GSD2022}.}  that (1) can provide a confidence region and (2) does not {\em a priori} require the {\iid} assumption. Our approach is inspired by the prediction-based ratio (PBR) protocol developed in~\cite{{ZGK11}} and the martingale-based method proposed by Gill~\cite{Gill2003,Gill2003b} for performing a hypothesis testing against the assumption of Bell-locality. Following~\cite{LZ19}, we further adapt these earlier methods and illustrate how they can be used for the device-independent certification of various properties of interest, including the underlying amount of entanglement and its fidelity with respect to some target quantum state.

To this end, we structure the rest of this paper as follows. In~\cref{Sec:Prelim}, we explain the basic concepts relevant to the understanding of DI certification in the ideal setting. After that, we introduce in~\cref{Sec:Finite} our adapted statistical tools for performing a rigorous device-independent certification. Results obtained from these tools are then presented in~\cref{Sec:Results}. Finally, we give some concluding remarks and future directions in~\cref{Sec:Discussion}.

\section{Preliminaries}\label{Sec:Prelim}

\subsection{Correlations and Bell inequalities}

The starting point of the DI approach is a Bell test. To this end, consider a bipartite Bell scenario, where two observers, Alice and Bob, can choose, respectively, their measurements labeled by $x,y \in \{0,1, ...\}$ and register outcomes  $a,b \in \{0,1, ...\}$.\footnote{If a third party is involved in the Bell test, as in the case of \cref{Sec:Prelim:ED,Sec:Results:ED}, we denote by $z$ and $c$, respectively, its label for the measurement setting and outcome. All other notations generalize accordingly.} In the {\em i.i.d.} setting, one can estimate the underlying correlation between measurement outcomes, i.e., $\vecP=\{P(ab|xy)\}$ from the registered empirical frequencies. Interestingly, as Bell first showed in~\cite{Bell64}, highly nontrivial conclusions can be drawn by inspecting $\vecP$ alone.

For example, correlations that can be produced in an LHV theory have to satisfy a Bell inequality:
\begin{equation}\label{Eq_BI}
	\sum_{x,y,a,b} \beta^{ab}_{xy} P(ab|xy) \overset{\L}{\le} B_{\L}(\vecB)
\end{equation}
where the {\em Bell coefficients} $\beta^{ab}_{xy}\in\mathbb{R}$, $\vecB:= \{\beta^{ab}_{xy}\}$, and $B_{\L}(\{\vecB\})$ is the so-called local (upper) bound. Here, we use $\L$ to signify that the inequality holds under the assumption that $\vecP$ is compatible with an LHV theory. Explicitly, the nature of such a theory demands that $\vecP$ is factorizable in the form of~\cite{Bell64,BCP+14}
\begin{equation}\label{Eq:BellLocal}
	P(ab|xy) \overset{\L}{=} \sum_\lambda q_\lambda P_A(a|x\lambda)P_B(b|y\lambda)
\end{equation}
where $q_\lambda\ge0$ for all $\lambda$, $\sum_\lambda q_\lambda =1$, and $P_A(a|x\lambda), P_B(b|y\lambda)\in[0,1]$ are local response functions.

In an actual Bell test, the measurement settings ought to be chosen randomly according to some predetermined distribution $P_{xy}$. To manifest this fact, 
one may write \cref{Eq_BI} using the unconditional joint distribution
 $P(abxy) = P(ab|xy)P_{xy}$ such that $P_{xy} = \sum_{a,b} P(abxy)$. In turn, we can then
 write a Bell inequality as a bound on the expectation value of a {\em Bell function} $I(v)$, defined in terms of $\vecB$ and $P_{xy}$, i.e.,
\begin{equation}\label{Eq_BI3}
	\left\langle I(v)\right\rangle := \tfrac{\left\langle\beta^{ab}_{xy}\right\rangle}{P_{xy}} \overset{\L}{\le} B_{\L}(\vecB)
\end{equation}
where $v=(a,b,x,y)$ is the quadruple of random variables for the measurement outcomes $(a,b)$ and settings $(x,y)$. As an example, the famous Clauser-Horne-Shimony-Holt (CHSH) Bell inequality \cite{{CHSH69}} may be specified via:
\begin{equation} \label{Bellfunction_CHSH}
    \ICHSH: \beta^{ab}_{xy}=(-1)^{xy+a+b}\quad\text{and}\quad B_{\L}=2,
\end{equation}
or equivalently, in terms of the correlator $E_{xy}:=\sum_{a,b=0,1} (-1)^{a+b} P(ab|xy)$, as:
\begin{equation} \label{Ineq_CHSH}
    \SCHSH=\sum_{x,y=0,1} (-1)^{xy}E_{xy}\overset{\L}{\le} 2,
\end{equation}
where $\SCHSH=\langle \ICHSH(v)\rangle$.

In contrast, quantum theory allows correlations that cannot be cast in the form of~\cref{Eq:BellLocal}. In fact, in a bipartite Bell test, general quantum correlations read as:
\begin{equation}\label{Eq:Born}
	P(ab|xy) \overset{\Q}{=} \tr(\rho\, M^{(A)}_{a|x}\otimes M^{(B)}_{b|y})
\end{equation}
where $\{M^{(A)}_{a|x}\}$ and $\{M^{(B)}_{b|y}\}$ are, respectively, the local positive-operator-valued measure (POVM) describing Alice and Bob's local measurements. For the benefits of subsequent discussions, it is also worth noting that both LHV and quantum correlations satisfy the nonsignaling conditions~\cite{PR94,BLM+05}:
\begin{equation}\label{Eq:NS}
\begin{split}
	\sum_a P(ab|xy) = \sum_a P(ab|x'y)\quad\forall\,\, x,x',\\
	\sum_b P(ab|xy) = \sum_b P(ab|xy')\quad\forall\,\, y,y'.
\end{split}
\end{equation}

For the CHSH Bell function, cf.~\cref{Bellfunction_CHSH}, quantum theory dictates the upper bound 
\begin{equation}\label{Eq_Tsirelson}
	\braket{\ICHSH(v)}\overset{\Q}{\le}B_{\Q}=2\sqrt{2},
\end{equation}
which can be seen as a Bell-like inequality. Other Bell and Bell-like inequalities relevant to this work will be presented in the corresponding sections below.

\subsection{Examples of properties to be certified}
\label{Sec:Properties}

\subsubsection{Negativity and dimension}
\label{Sec:NegDim}

As mentioned above, with local measurements on a quantum system, a Bell-inequality-violating correlation $\vecP\not\in\L$ necessarily originates~\cite{Werner:PRA:1989} from an entangled state $\rho$. Interestingly, the entanglement of the underlying $\rho$  can also be lower bounded~\cite{MBL+13,TMG15,CBLC18,AB19} directly from the observed correlation~$\vecP$. In this work, we focus on negativity~\cite{Vidal02} but it is worth noting that DI entanglement quantification can be also achieved, e.g., for the linear entropy of entanglement~\cite{TMG15}, generalized robustness of entanglement~\cite{CBLC18}, and one-shot distillable entanglement~\cite{AB19}.

For a bipartite density operator $\rho$, let $\rho^{\mbox{\tiny T}_A}$ be its partial transposition~\cite{Peres96} with respect to subsystem $A$. Then, the negativity for a bipartite density operator $\rho$ is defined as~\cite{Vidal02} $\Neg(\rho):=\sum_{\lambda_i<0} |\lambda_i(\rho^{\Tp_A})|$, i.e., the sum of the absolute value of all negative eigenvalues $\lambda_i<0$ of $\rho^{\mbox{\tiny T}_A}$. Using a variational characterization of negativity provided in~\cite{Vidal02}, it was shown in~\cite{MBL+13} that $\Neg(\rho)$ is lower bounded by the optimum value of the following semidefinite program (SDP):
\begin{subequations}\label{Eq:SDP:MinNeg}
\begin{align}
     \text{min\ \ \ } &\chi_\ell[\sigma_-]_{\text{tr}} \\
    \text{s.t. \ \ } & \chi_\ell [\rho] = \chi_\ell[\sigma_+]-\chi_\ell[\sigma_-] , \quad \chi_\ell[\sigma_\pm]^{\Tp_{\bar{A}}} \succeq 0, 
    \label{Eq:SDP:MinNeg-Constraints}\\
     & \chi_\ell [\rho]\succeq 0,\quad \chi_\ell[\rho]_{\text{tr}} = 1,      \label{Eq:SDP:Quantum}
\end{align}
\end{subequations}
where $\chi_\ell[\rho]$ is a moment matrix that can be obtained by applying a particular local map on $\rho$ (see~\cite{MBL+13} for details), $\bar{A}$ is the output Hilbert space of the local map on $A$, $\chi_\ell[ \sigma ]_{\text{tr}} = \text{tr} [\sigma]$ represents the trace of the underlying operator $\sigma$.
It is worth noting that for every integer $\ell\ge 1$, the constraints of \cref{Eq:SDP:Quantum} provide a superset characterization of the quantum set $\Q$ of correlations, analogous to those considered in~\cite{NPA,NPA2008,DLTW08}. Indeed, all entries from $\vecP$ appear somewhere in the moment matrix $\chi_\ell[\rho]$, see~\cite{MBL+13}.

As an explicit example, note that an observed {\em violation} of the CHSH Bell inequality of \cref{Ineq_CHSH} gives the following nontrivial negativity lower bound of the underlying state $\rho$:
\begin{equation}\label{Eq:CHSH_Neg}
	\Neg(\rho)\ge \frac{\SCHSH-2}{4(\sqrt{2}-1)}.
\end{equation}

Also worth noting is that if $\rho$ acts on $\mathbb{C}^{d_A} \otimes \mathbb{C}^{d_B}$ with $d=\min\{d_A,d_B\}$, then the maximal possible negativity $\Neg(\rho)$ is upper bounded by $\Neg^d_\text{max}:=\frac{d-1}{2}$. Consequently, the observation of a large enough negativity also provides a nontrivial lower bound on the local Hilbert space dimension of the underlying system. More precisely, if the lower bound on $\Neg(\rho)$ obtained from \cref{Eq:SDP:MinNeg} exceeds $\Neg^d_\text{max}$, one immediately deduces that $\rho$ must act on a local Hilbert space of dimension $\ge d+1$, thereby giving a dimension witness~\cite{BPA+08}. 

From \cref{Ineq_CHSH,Eq_Tsirelson,Eq:CHSH_Neg}, nonetheless, we see that a violation of the CHSH Bell inequality can never witness a local Hilbert space dimension $>2$. Instead, witnessing a local Hilbert space beyond qubits can be achieved by observing a reasonably strong violation of the 3-outcome Collins-Gisin-Linden-Massar-Popescu (CGLMP) Bell inequality~\cite{CGLMP02} (see also~\cite{KKC+02}), defined by
\begin{align}
	\ICGLMP: \beta^{ab}_{xy} = &(-1)^{x(y-1)}\{\delta^{(2)}_{a-b}-[1-\delta^{(2)}_x\delta^{(2)}_{y-1}]\delta^{(3)}_{b-a-1}\}\nonumber\\
	&-\delta^{(2)}_x\delta^{(2)}_{y-1}\delta^{(3)}_{b-a+1}\quad\text{and}\quad B_{\L}(\vecB)=2,\label{Ineq:CGLMP}
\end{align}
where $\delta^{(d)}_f=1$ if $\text{mod}(f,d)=0$ and vanishes otherwise. Denoting the corresponding expectation value by $\SCGLMP=\langle\ICGLMP(v)\rangle$, the results from~\cite{Acin2002,MBL+13,Liang:PhDthesis} suggest a negativity lower bound that increases linearly with $\SCGLMP$ from $\frac{1}{2}$ whenever $\SCGLMP\ge \frac{3}{\sqrt{2}}+\frac{1}{2}$.

\subsubsection{Entanglement depth}
\label{Sec:Prelim:ED}

In a many-body system, entanglement can come in various forms or structures~\cite{HZL+18}. In particular, an $n$-partite quantum state that is not fully separable is {\em not} necessarily genuinely $n$-partite entangled either. To witness the latter, one could rely on the demonstration of so-called genuine multipartite nonlocality~\cite{BBG+13}. However, as remarked in~\cite{BGLP11}, it is possible to witness genuine multipartite entanglement without relying on this strong form of multipartite nonlocality. In fact, using the SDP  introduced in~\cite{MBL+13}, one can even systematically construct DI witnesses of this kind, starting from a given multipartite Bell function, say $\vecB=\{\beta^{abc}_{xyz}\}$. Later, it was further shown in~\cite{LRB+15} (see also~\cite{Curchod15}) that the extent to which a multipartite Bell inequality is violated can be used to witness (lower-bound) the underlying entanglement depth~\cite{Guhne:NJP:2005,Sorensen2001May}, i.e., the extent to which a many-body entanglement is needed to prepare the given multipartite state. 

For illustration, consider the expectation value of the Mermin Bell function~\cite{Mermin:PRL:1990} $I_\text{\tiny Mermin}(v)$ with $v=(a,b,c,x,y,z)$:
\begin{equation}\label{Ineq:Mermin}
	\SMermin=\ExpVal{I_\text{\tiny Mermin}(v)} =\frac{\ExpVal{\beta^{abc}_{xyz}}}{P_{xyz}} = \sum_{x,y,z}{}^{'} (-1)^{xyz} E_{xyz}
\end{equation}
where $E_{xyz}:=\sum_{a,b,c=0}^1 (-1)^{a+b+c}P(abc|xyz)$ is the tripartite correlator, the restricted sum $\sum'$ is over all combinations of $x,y,z\in\{0,1\}$ such that $\text{mod}(x+y+z,2)=1$, $P_{xyz}=\frac{1}{4}$ for the same combinations of $x,y,z$, and the Bell coefficients are
\begin{equation}
	\beta^{abc}_{xyz} = (-1)^{xyz+a+b+c}\delta^{(2)}_{x+y+z-1}.
\end{equation}
Then, it is known that the following Bell-like inequalities hold, respectively, for fully-separable states, 2-producible~\cite{Guhne:NJP:2005} tripartite quantum states (i.e., quantum states that can be generated using only 2-body entanglement), and general tripartite quantum states:
\begin{equation}\label{Eq:Mermin:Bounds}
	\SMermin\overset{\L}{\le} 2,\quad \SMermin\overset{\text{2-prod.}}{\le} 2\sqrt{2},\quad \SMermin\overset{\Q}{\le} 4.
\end{equation}

\subsubsection{State fidelity}

The strongest form of device certification one can hope for within a DI paradigm is called {\em self-testing}~\cite{SB20}, first proposed in~\cite{MY98}. The key observation behind this feat is that the quantum strategy compatible with {\em certain} extremal quantum correlations $\vecP_\Q$ is essentially unique. Hence, with the observation of $\vecP_\Q$ in a Bell test, we can conclude, unambiguously that some degree of freedom (DOF) of the measured system  {\em must} match a specific target state $\ket{\psi}$. Often, one can also self-test the underlying measurements alongside the state (see, however,~\cite{Jeba:PRR:2019,Kaniewski:PRR:2020} for some examples of exceptions).

For instance, it is long known~\cite{Summers:1987aa,Popescu:1992aa,Braunstein1992,Tsirelson93} that the maximal CHSH Bell-inequality violation of $\SCHSH=2\sqrt{2}$ can only obtained (up to local isometry) by measuring the following  observables on a shared maximally entangled state (MES):
\begin{subequations}\label{Eq:CHSH:Strategy}
\begin{gather}
\ket{\psi_\text{MES}}=\frac{1}{\sqrt{2}}(\ket{00}+\ket{11}),\label{Eq:MES2}\\
A_0 = \sigma_z,\quad A_1=\sigma_x,\\
B_y = \frac{1}{\sqrt{2}}[\sigma_z + (-1)^y\sigma_x],\label{Eq:OptMeas:CHSH}
\end{gather}
\end{subequations}
where the respective POVM elements (with $x,y=0,1$) are
\begin{equation}
	M^{(A)}_{a|x}=\frac{\Id+(-1)^aA_x}{2},\quad M^{(B)}_{b|y}=\frac{\Id+(-1)^bB_y}{2}.
\end{equation}

Moreover, to obtain the maximal CHSH Bell-inequality violation for a partially entangled two-qubit state,
\begin{subequations}\label{Eq:Sefl-test:PES}
\begin{equation}\label{Eq:PsiTheta}
	\ket{\psi(\theta)}=\cos\theta\ket{00}+\sin\theta\ket{11},\quad \theta\in\left(0,\frac{\pi}{4}\right]
\end{equation}
it suffices~\cite{Liang:PhDthesis} to consider $A_x$ of \cref{Eq:OptMeas:CHSH} but generalize $B_y$ to~\cite{Bancal15}:
\begin{gather}
B_y = \cos\mu\,\sigma_z +(-1)^y \sin\mu\,\sigma_x,\quad \tan\mu = \sin(2\theta),
\end{gather}
\end{subequations}
thereby giving 
\begin{equation}\label{Eq:CHSH:General}
	\SCHSH=2\sqrt{1+\sin^22\theta}. 
\end{equation}
Interestingly, the resulting correlation also self-tests~\cite{Yang13,Bancal15} the corresponding quantum strategy of \cref{Eq:Sefl-test:PES} and maximally violate the family of tilted CHSH Bell inequalities for $\alpha=2\sqrt{\frac{\cos^22\theta}{1+\sin^22\theta}}$:
\begin{equation}\label{Ineq:TCHSH}
	\S^\text{\tiny Tilted}_\text{\tiny CHSH}(\alpha) = \SCHSH + \alpha\sum_{a, b=0}^1 (-1)^aP(ab|0y) \overset{\L}{\le} 2+\alpha,
\end{equation}
giving $\S^\text{\tiny Tilted}_\text{\tiny CHSH}(\alpha)= \sqrt{8+2\alpha^2}$. Note that in \cref{Ineq:TCHSH}, thanks to the nonsignaling~\cite{PR94,BLM+05} property of $\vecP$, the expression for $\S^\text{\tiny Tilted}_\text{\tiny CHSH}(\alpha)$ is in fact independent of whether $y=0$ or $1$.

In practice, however, due to various imperfections, one can, at best attain a correlation close to the ideal one $\vecP_\Q$. In other words, in a realistic experimental setting, one can only hope to 
lower bound the similarity of the measured state $\rho$ with respect to the target state $\ket{\psi}$ via a fidelity measure. To this end, a powerful numerical technique known as the SWAP method has been introduced in~\cite{YVB+14} (see also~\cite{Bancal15}) for exactly this purpose. More precisely, for any observed quantum correlation $\vecP$, the method allows one to lower bound the fidelity:
\begin{equation}
	\F = \bra{\psi}\rhoS\ket{\psi}
\end{equation}
with the help of an SDP outer approximation of the quantum set $\Q$ (e.g., due to~\cite{NPA,NPA2008,MBL+13}). Here, $\rhoS$ is the ``swapped" state:
\begin{equation}
	\rhoS = \tr_{AB}[\Phi\,\rho_{AB}\otimes (\proj{00})_{A'B'}\,\Phi^+]
\end{equation}
extracted from the underlying quantum state $\rho$ via some local extraction map $\Phi$, which is a function of the actual POVM elements. Consequently, $\F$ is a function of the entries of the moment matrix $\chi_\ell[\rho]$, discussed below \cref{Eq:SDP:MinNeg}.  For details of the method, we refer the readers to~\cite{Bancal15}.

\subsection{Some general remarks}
\label{Sec:Remarks}

At this point, it is worth noting that for all the three properties $\P$ discussed above---negativity (and hence dimension), entanglement depth, and reference-state fidelity---their DI certification can be achieved via the characterization of some {\em convex set} $\mathcal{C}_\P$ in the space of correlation vectors $\{\vecP\}$. More precisely, for negativity, by turning the objective function of \cref{Eq:SDP:MinNeg} into the constraint~\cite{MBL+13} 
\begin{equation}
	\chi_\ell[\sigma_-]_{\text{tr}}\le \Neg_0,
\end{equation}
we obtain an SDP that characterizes of the set of correlations attainable by quantum states having a negativity upper-bounded by $\Neg_0$. Then, \cref{Eq:CHSH_Neg} can be understood as a separating hyperplane relevant for witnessing a negativity larger than $\Neg_0$.

On the other hand, if we drop the constraint in \cref{Eq:SDP:MinNeg} but imposes additional positive-partial-transposition constraints, then we get an SDP characterization of the set $\mathcal{C}_\P$ having a bounded amount of entanglement depth~\cite{LRB+15} [see constraints of \cref{Eq:PBR:Bisep:PStar} below]. In this case, the first two inequalities of \cref{Eq:Mermin:Bounds} serve as the corresponding witness for entanglement depth. 
Finally, by demanding $\bra{\psi}\rhoS\ket{\psi}\le\F_0$ together with \cref{Eq:SDP:Quantum}, we obtain an SDP characterization of the set $\mathcal{C}_\P$ associated with a swapped state~\cite{Bancal15} with a $\ket{\psi}$-fidelity upper bounded by $\F_0$. In fact, an SDP characterization can also be obtained for a number of other properties, including genuine negativity~\cite{MBL+13}, steering robustness~\cite{CBLC16}, entanglement robustness~\cite{CBLC18}, (measurement) incompatibility robustness~\cite{CBLC16,CMBC21}, and so on.

\section{Methodologies for hypothesis testing}\label{Sec:Finite}

Having understood how DI certification can be achieved from a given correlation $\vecP$, we now proceed to discuss the more realistic setting involving only a finite number of experimental trials. For concreteness, the following presentation assumes an analysis based on the data collected from $N$ trials in a Bell test. Below, we explain our approaches to the problem based on {\em hypothesis testing}. Our first step is to formulate a {\em null hypothesis} $\H$ based on the desired property to be certified. For example, to certify that the underlying state has a negativity larger than $\Neg_0$, we formula the (converse) null hypothesis: 

\begin{hypothesis}\label{H:Negativity}
    $\H_{\Neg(\rho)\le \Neg_0}$: In every experimental trial, the underlying state has a negativity less than or equal to $\Neg_0$. 
\end{hypothesis}
Since such a hypothesis involves a {\em set} of (rather than a single) compatible distribution $\vecP$, it is called a {\em composite hypothesis}~\cite{vDGG05}.

Then, we apply appropriate methods for this kind of hypothesis testing on the collected data to determine an upper bound $\mathfrak{p}$ on the $p$-value associated with the hypothesis $\H$. Since a $p$-value quantifies the plausibility of observing the given data when $\H$ holds, a small value of $\mathfrak{p}$, say, less than $5\%$ provides a strong indication that $\H$ is falsified. It then follows that the desired feature corresponding to the negation of $\H$ is certified with a confidence $\gamma$ of at least $1-\mathfrak{p}$.

Of course, one may also be interested to understand how quickly statistical evidence (against a hypothesis $\H$) can be gathered when we increase the number of trials. To this end, we also consider the so-called (asymptotic) confidence-gain rate~\cite{ZGK11}, defined by 
\begin{equation}\label{confidencegainrate}
    G\prot := - \underset{\Ntot \rightarrow \infty}{\text{lim}} \frac{\log_2 p\prot_{\Ntot}}{\Ntot}, 
\end{equation}
where $p^{\text{(prot)}}$ is the $p$-value (upper bound) deduced from some protocol (abbreviated as ``prot"). From the definition, it is evident that asymptotically, and in the {\em i.i.d.} setting, a fewer number of trials is required to achieve the same level of statistical confidence if the corresponding $G\prot$ is higher. Next, let us elaborate on the two hypothesis-testing protocols considered in this work.

\subsection{Martingale-based protocol}
\label{Sec. Martingale-based protocol}

We shall start with the martingale-based protocol, pioneered by Gill in~\cite{Gill2003,Gill2003b} for testing against LHV theories, and further developed in~\cite{ZGK11,ZGK13}. The protocol relies on the observation of the (super)martingale structure in some random variable of interest. To employ the martingale-based protocol, one has to fix a Bell function $I(v)$ in advance. Ideally, $I(v)$ should be chosen such that the Bell-like inequality
\begin{equation}\label{Ineq:Bell-like:H}
    \ExpVal{I(v)}:=
    \ExpVal{\tfrac{\beta^{ab}_{xy}}{P_{xy}} }=\sum_{a,b,x,y} \beta^{ab}_{xy} P(ab|xy)\overset{\H}{\le} B_\H(\vec{\beta})
\end{equation}
may be violated by some quantum correlation $\vecP=\vecP_\Q$ [cf.~\cref{Eq:Born}] to be prepared in an experiment.

Let $v_j=(a_j,b_j,x_j,y_j)$ be the value realized for the random variables of the measurement outcomes and settings at the $j$-th experimental trial, and $I(v_j)$ the corresponding value of Bell function for that trial. Moreover, let $\bfv=\{v_1,...,v_i,..., v_{N}\}$. Then, from the observed average value of $I(v)$ over $N$ trials, i.e., $\hat{I}(\bfv) = \sum_{j=1}^{N} \frac{I(v_j)}{N}$, the following $p$-value upper bound is known~\cite{ZGK13} to hold whenever $\hat{I} \ge B_\H$:
\begin{align}\label{Eq:mart2}
    p\mart \leq \left[ \left(\frac{\mathfrak{b}_+-B_\mathcal{H}}{\mathfrak{b}_+-\hat{I}} \right)^{\frac{\mathfrak{b}_+-\hat{I}}{\mathfrak{b}_+-\mathfrak{b}_-}} \left(\frac{B_\mathcal{H}-\mathfrak{b}_-}{\hat{I}-\mathfrak{b}_-}\right)^{\frac{\hat{I}-\mathfrak{b}_-}{\mathfrak{b}_+-\mathfrak{b}_-}} \right]^N,
\end{align}
where, for simplicity, we have suppressed the dependency of $B_\H$ on $\vecB$ (and $\hat{I}$ on $\bfv$), while the minimum and maximum value of $I(v)$ over all possible values of $v=(a,b,x,y)$ are
\begin{equation}\label{Eq:bpm}
	\mathfrak{b}_-:= \inf_v I(v) < B_\H \le \hat{I} < \mathfrak{b}_+:= \sup_v I(v).
\end{equation}
It is worth noting that the martingale-based $p$-value upper bound of \cref{Eq:mart2} improves over the one given in~\cite{Gill2003b,PMG+10,ZGK11}.

\begin{figure}[H]
\begin{mdframed}

{\bf Inputs:} Property to be certified $\P$, raw data $\bfv =\{v_i=(a_i,b_i,x_i,y_i)\}_{i=1}^{\Ntot}$ acquired in a Bell test, a preset confidence level $\gamma$ (we use $\gamma=99\%$ in all the examples below), a Bell-like inequality, cf. \cref{Ineq:Bell-like:H}, compatible with the Null Hypothesis $\H_{\neg\P}$ (associated with the {\em negation} of $\P$).\\
{\bf Output:} A $p$-value bound $\mathfrak{p}$ for the Null Hypothesis $\H_{\neg\P}$ to be compatible with $\bfv$. \\
{\bf Steps:}
\begin{enumerate}
	\item Compute $\mathfrak{b}_\pm$ from \cref{Eq:bpm} and the average value $\hat{I}(\bfv) = \frac{1}{\Ntot}\sum_{i=1}^{\Ntot} I(v_i)= \frac{1}{\Ntot}\sum_{i=1}^{\Ntot} \tfrac{\beta^{a_ib_i}_{x_iy_i}}{P_{x_iy_i}}$.
	\item Compute the required $p$-value bound $\mathfrak{p}$ by substituting the determined $\mathfrak{b}_\pm$ and $\hat{I}(\bfv)$ into \cref{Eq:mart2} with $N$ replaced by $\Ntot$.
\end{enumerate}

\end{mdframed}
\caption{Pseudocode associated with the martingale-based protocol for DI certification. If the $p$-value bound $\mathfrak{p}<1-\gamma$, we reject the Null Hypothesis $\H_{\neg\P}$ and hence certify the desired property $\P$ with confidence $\gamma$.}
\end{figure}

Let $I_\Q$ be the expectation value of $I(v)$ when we replace $\vecP$ by some $\vecP_\Q$ capable of violating the Bell-like inequality in \cref{Ineq:Bell-like:H}. Then, in the {\em i.i.d.} setting where the experimental data follows the distributions given by $\vecP_\Q$, the corresponding asymptotic confidence-gain rate can be deduced from \cref{confidencegainrate} and \cref{Eq:mart2} as:
\begin{equation}\label{Eq:Mart:G}
    G\mart = \frac{\mathfrak{b}_+-I_\Q}{\mathfrak{b}_+-\mathfrak{b}_-}\log_2 \frac{\mathfrak{b}_+-I_\Q}{\mathfrak{b}_+-B_{\mathcal{H}}} + \frac{I_\Q-\mathfrak{b}_-}{\mathfrak{b}_+-\mathfrak{b}_-} \log_2 \frac{I_\Q-\mathfrak{b}_-}{B_{\mathcal{H}}-\mathfrak{b}_-}.
\end{equation}

\subsection{The prediction-based-ratio (PBR) protocol}
\label{Sec:PBR}

The other hypothesis-testing protocol that we consider in this work is based on the so-called PBR protocol proposed in~\cite{ZGK11} (see also~\cite{ZGK13}). In contrast with a martingale-based protocol, the PBR protocol does {\em not} need to presuppose any Bell-like inequality for determining a $p$-value bound. Instead, for the data $\bfv$ collected in $N$ trials, one may start by using the first $\Nest< N$ trials from $i=1,2,\cdots,\Nest$ to estimate the relative frequency
\begin{align}\label{Eq:f}
    f(ab|xy) = \frac{\Nest(a,b,x,y)}{\Nest(x,y)},
\end{align}
where $\Nest(x,y)=\sum_{a,b} \Nest(a,b,x,y)$ and $\Nest(a,b,x,y)$ counts among these $\Nest$ trials the total number of times the {\em specific} combination of measurement settings and outcomes $(x,y,a,b)$ occurs.

The key idea of the PBR protocol is to use this relative frequency $\vecf=\{f(ab|xy)\}$ to obtain an {\em optimized Bell-like inequality}\footnote{Here, the inequality is optimized in the sense that it provides the largest possible asymptotic confidence-gain rate, cf.~\cref{confidencegainrate}.} and apply that to $v_i$ from $i=\{\Nest+1,\Nest+2,\cdots,\Nest+\Ntest\}$. To this end, we minimize the Kullback-Leibler (KL) divergence \cite{Kullback1951Mar} from a regularized relative frequency $\vecf\reg$ (explained below) to the set $\S_\H$ of correlations compatible with $\H$:
\begin{align}
    \displaystyle D_{\text{KL}} & (\vecf\reg||\S_\H) := \nonumber \\
     & \inf_{\vecP \in \S_\H} \sum_{a,b,x,y} P_{xy} f\reg(ab|xy) \log \frac{f\reg(ab|xy)}{P(ab|xy)}. \label{Eq:KLD}
\end{align}
An important point to note now is that if the composite null hypothesis $\S_\H$ is associated with a convex set that admits an SDP characterization (as discussed in~\cref{Sec:Remarks}) like the kind proposed in~\cite{NPA,NPA2008,DLTW08,MBL+13,CBLC16}, then \cref{Eq:KLD} is a conic program (see~\cite{LRZ+18}), and thus efficiently solvable using a solver like $\mathsf{MOSEK}$~\cite{Mosek_exponential_cone}.

The unique~\cite{LRZ+18} minimizer $\vecP_\star\in\S_\H$ can then be used to define the {\em non-negative} prediction-based-ratio (PBR)
\begin{align}\label{Eq:PBR}
    R(a,b,x,y):= \frac{f\reg(ab|xy)}{P_\star(ab|xy)},
\end{align}
which gives the optimized Bell-like inequality $\ExpVal{R(v)}\overset{\H}{\le} 1$.
Next, we compute the test statistic 
\begin{equation}
	t(\bfv)=\prod_{j} R(a_j,b_j,x_j,y_j)
\end{equation}
where the product is only carried out over the remaining $\Ntest$ trials.
Using arguments completely analogous to those given in~\cite{ZGK11} for $\H=\L$, it can then be shown that the following upper bound on the $p$-value holds:
\begin{align}\label{Eq:p-PBR}
    p\pbr \leq \text{min} \left\{ \frac{1}{t(\bfv)},1 \right\}.
\end{align}

Several remarks are now in order. Firstly, if none of the entries in $\vecf$ vanishes, one could also use $\vecf$ directly in the optimization problem of \cref{Eq:KLD}. However, for a small $\Nest$, a vanishing entry in $\vecf$ is almost bound to happen, we thus follow~\cite{ZGK11} and mix $\vecf$ with the {\em uniform distribution} $\vecP_\Id$ to obtain:
\begin{equation}\label{Eq:mixedF}
	\vecf \to \vecf':= \frac{\Nest}{\Nest+1}\vecf + \frac{1}{\Nest+1}\vecP_\Id.
\end{equation}
Next, notice that $\vecf'$ typically cannot be cast in the form of \cref{Eq:Born}. Consequently, we observe empirically that the $R$ obtained by solving \cref{Eq:KLD} with $\vecf'$ in place of $\vecf\reg$ gives evidently suboptimal performance (see, e.g., \cref{fig:CHSH_lv3_ga099_fourmethods,fig:CGLMP_lv14_ga099_100k_fourmethods,fig:CHSH_extractability_ga099_fourmethods} in~\cref{App:MISC} for some explicit examples). As such, we shall first regularize~\cite{LRZ+18} $\vecf'$ to some outer approximation of the quantum set $\Q_\ell$ by solving \cref{Eq:KLD} with $\S_\H$ replaced by $\Q_\ell$. In our work, $\Q_\ell$ is the level-$\ell$ outer approximation of the quantum set $\Q$ introduced in \cite{MBL+13}. However, one may also consider other approximations~\cite{NPA,CBLC16}. Since all these outer approximations admit an SDP characterization, this regularization process is a conic program (see also~\cite{LRZ+18}). The resulting minimizer, which we call the regularized relative frequency $\vecf\reg$ is then fed into \cref{Eq:KLD} to obtain the desired PBR.

\begin{figure}[H]
    \centering
    \includegraphics[width=0.5\textwidth]{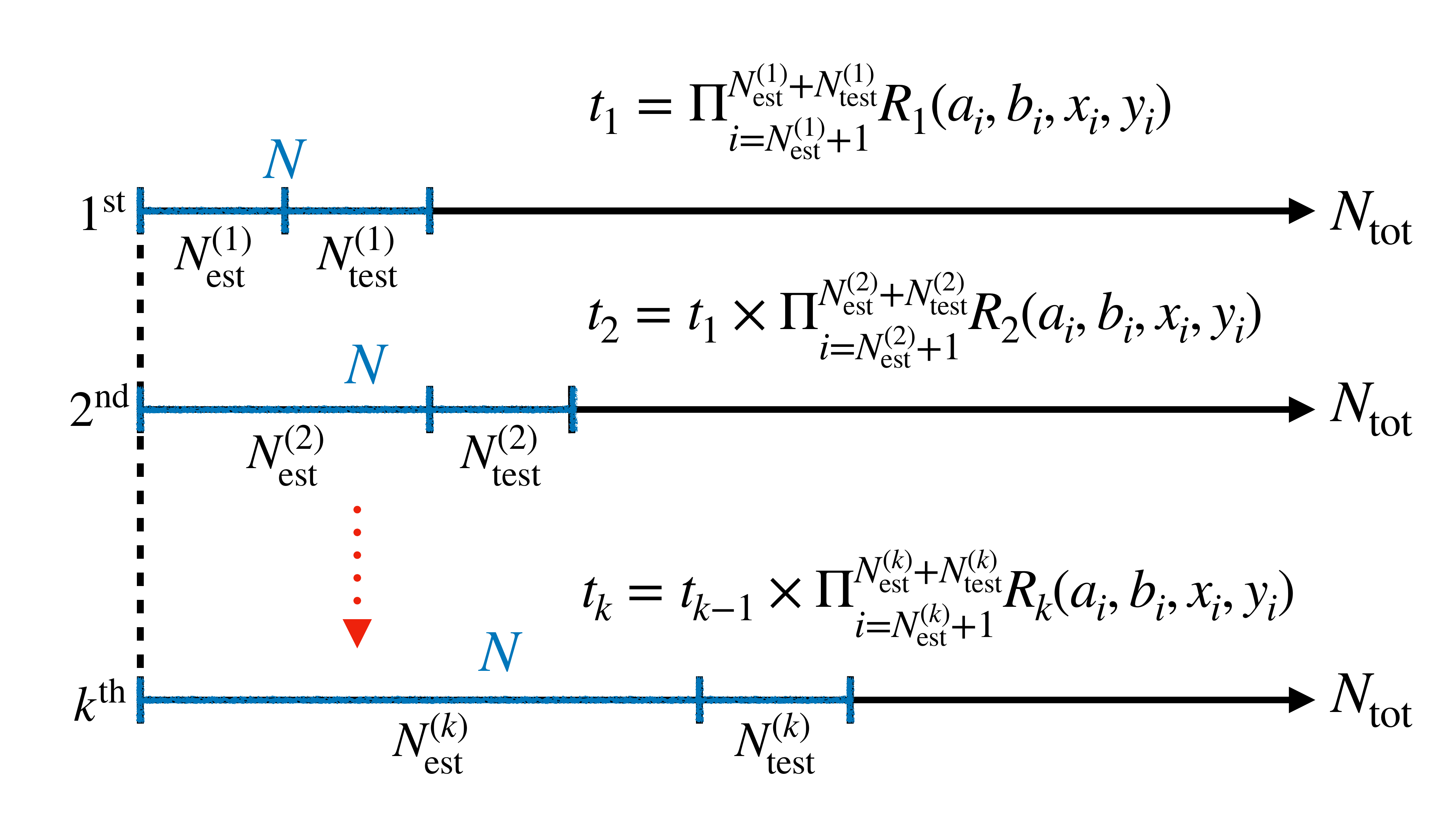}
    \caption{Instruction for the PBR method.}
    \label{pbr_instruction}
\end{figure}

Another important feature of the PBR protocol is that the optimized inequality characterized by $\vec{R} = \{R(a,b,x,y)\}$ can be updated as more data is incorporated into the analysis. In principle, one can update $\vec{R}$ as frequently as one desires. However, this is neither necessary nor efficient. As such, we work with blocks of $\Nint$ trials. The first block of data is used exclusively for producing the first regularized relative frequency, the first PBR $\vec{R}_1$, and by applying to the second block of $\bfv$, we get the first test statistic
\begin{equation}\label{Eq:t1}
    t_1=\Pi^{N^{(1)}_{\text{est}}+N^{(1)}_{\text{test}}}_{i=N^{(1)}_{\text{est}}+1} R_1 (a_i,b_i,x_i,y_i),
\end{equation}
where $N^{(k)}_{\text{test}} = \Nint$ for all $k$. In the next iteration, we determine the PBR $\vec{R}_2$ by solving \cref{Eq:KLD} using $\bfv$ from the first two blocks, and apply this updated PBR to the third block of $\bfv$ to get, for $k=2$,
\begin{equation}\label{Eq:tk}
	t_k = t_{k-1} \times \Pi^{N^{(k)}_{\text{est}}+N^{(k)}_{\text{test}}}_{i=N^{(k)}_{\text{est}}+1} R_k (a_i,b_i,x_i,y_i)
\end{equation}
where $N^{(k)}_{\text{est}}=k\Nint$. These steps may then repeated iteratively until all the data $\bfv$ has been consumed in one way or another in the computation of $t_k$ for 
$k=3,\ldots,\frac{\Ntot}{\Nint}-1$. For a schematic illustration of this procedure, see~\cref{pbr_instruction}. Importantly, once the test statistic $t_k$ for each iteration is determined, we can obtain the corresponding $p$-value bound using~\cref{Eq:p-PBR}.

\begin{figure}[H]
\begin{mdframed}
{\bf Inputs:} Property to be certified $\P$, raw data $\bfv =\{v_i=(a_i,b_i,x_i,y_i)\}_{i=1}^{\Ntot}$ acquired in a Bell test, a preset confidence level $\gamma$ (we use $\gamma=99\%$ in all the examples below), and the distribution $\{P_{xy}\}_{x,y}$.\\
{\bf Output:} A $p$-value bound $\mathfrak{p}$ for the Null Hypothesis $\H_{\neg\P}$ (associated with the negation of $\P$) to be compatible with $\bfv$. \\
{\bf Steps:}
\begin{enumerate}
	\item Choose an appropriate block size $\Nint$ (we use $\Nint=500$ in the examples below).
	\item Set $k=1$ and $t_0=1$.
	\item\label{Step:vecf} Use the block of data $\bfv =\{v_i=(a_i,b_i,x_i,y_i)\}_{i=1}^{N^{(k)}_\text{est}}$ to compute the relative frequency $\vecf$ via \cref{Eq:f}, where $N^{(k)}_{\text{est}}=k\Nint$ if the trials are expected to be (near) {\em i.i.d.}
	\item If $\vecf$ contains zero(s), apply \cref{Eq:mixedF} to obtain $\vecf\,'$ with only nonvanishing entries; else, set $\vecf\,'=\vecf$.
	\item Solve the optimization of \cref{Eq:KLD} with $\vecf\,'$ playing the role of $\vecf\reg$ and $\S_\H$ replaced by $\Q_\ell$ [in most examples below, we use the level $\ell=3$ approximation of $\Q$ introduced in~\cite{MBL+13} for $\Q_\ell$]. 
	\item Solve the optimization of \cref{Eq:KLD} using the optimizer from the last step as $\vecf\reg$ where $\S_\H$ is now the set of correlations compatible with $\H_{\neg\P}$.
	\item Use $\vecf\reg$ and the resulting optimizer $\vecP_\star$ in \cref{Eq:PBR} to determine the prediction-based-ratio for the $k$-th iteration, $\{R_k(a,b,x,y)\}_{a,b,x,y}$.
	\item Use $R_k$ and the next block of data $\bfv =\{v_i=(a_i,b_i,x_i,y_i)\}_{i=N^{(k)}_\text{est}+1}^{N^{(k)}_\text{est}+N^{(k)}_\text{test}}$ in \cref{Eq:tk} to determine the test statistic $t_k$ for the $k$-th iteration. Here, $N^{(k)}_\text{test}=\min\{\Nint, \Ntot -N^{(k)}_\text{est} \}$.
	\item\label{Step:quit} Unless all $v_i$ have been used in the computation, increase $k$ by 1.
	\item Repeat steps \ref{Step:vecf} to \ref{Step:quit} until all $v_i$ have been used in the computation.
	\item Compute the required $p$-value bound $\mathfrak{p}$ by using the last $t_k$ computed as $t(\bfv)$ in \cref{Eq:p-PBR}.
\end{enumerate}
\end{mdframed}
\caption{Pseudocode associated with the PBR protocol for DI certification. If the $p$-value bound $\mathfrak{p}<1-\gamma$, we reject the Null Hypothesis $\H_{\neg\P}$ and hence certify the desired property $\P$ with confidence $\gamma$.}
\end{figure}

Finally, note that for an ideal Bell test giving the correlation $\vecP_\Q$ and a composite hypothesis associated with $\H$, the PBR protocol has the asymptotic confidence-gain rate:
\begin{equation}\label{Eq:PBR:G}
	G\pbr = D_{\text{KL}}(\vecP_\Q||\S_\H),
\end{equation}
which may be obtained by solving \cref{Eq:KLD} with $\vecf\reg$ replaced by $\vecP_\Q$. The proof is again completely analogous to that given for $\H=\L$ in~\cite{ZGK11} and is thus omitted.

\section{Device-independent certification with a confidence interval}\label{Sec:Results}

We are now ready to present our simulations results involving a finite number of trials. Throughout this section, the results presented for finite trials consist of an average over $30$ complete Bell tests, each involving $\Ntot=10^5$ trials, with the trials partitioned into blocks of size $\Nint=500$.
Moreover, we always consider a uniform distribution for (possibly a restricted set of) measurement settings.
In each Bell test, we then simulate the raw data $\bfv =\{v_i=(a_i,b_i,x_i,y_i)\}_{i=1}^{\Ntot}$ using the function $\mathsf{sample\_hist}$ from the Lightspeed Matlab toolbox~\cite{lightspeed}. For the certification with finite data, we set a confidence level of $\gamma=0.99$. We also present some related confidence-gain rates in the respective subsections.

\subsection{Negativity and dimension certification}
\subsubsection{Negativity certification}\label{Subsec. CHSH}

Our first example consists of a Bell test based on the quantum strategy presented in \cref{Eq:CHSH:Strategy}, which leads to a CHSH Bell value of $\SCHSH=2\sqrt{2}$. Using \cref{Eq:CHSH_Neg}, we know that the resulting quantum correlation $\PCHSH$ gives a tight negativity lower bound of $\frac{1}{2}$ for a Bell state. From the numerically simulated data, 
we then perform composite hypothesis testing for  Null Hypothesis~\ref{H:Negativity} with $\Neg_0 \in \{ 0,0.01,\cdots,0.50 \}$.

Specifically, for the martingale-based protocol, we use \cref{Eq:mart2} with the CHSH Bell expression of \cref{Bellfunction_CHSH}. In this case, $\mathfrak{b}_\pm=\pm4$ for the chosen $P_{xy}$ while it follows from \cref{Bellfunction_CHSH,Eq:CHSH_Neg,Ineq:Bell-like:H} that 
\begin{equation}\label{Eq:Mart:CHSH:Neg}
	\sum_{a,b,x,y} (-1)^{xy+a+b} 4\,P(abxy) \overset{\Neg\le \Neg_0}{\le} 2+4\Neg_0(\sqrt{2}-1).
\end{equation} 
On the other hand, for the PBR protocol, the optimizing distribution $P^{(k)}_\star(a,b|x,y)$ for the $k$-iteration can be obtained by solving [cf. \cref{Eq:KLD}]
\begin{subequations}\label{Eq:PBR:Neg:PStar}
\begin{align}
      \argmin_{\vecP} \,\,&-\sum_{a,b,x,y} P_{xy} f\reg^{(k)}(ab|xy) \log P(ab|xy),\label{Eq:Argmin}\\
    \text{s.t. \ \ }  \chi_\ell [&\rho] = \chi_\ell[\sigma_+]-\chi_\ell[\sigma_-] \succeq 0, \,\, \chi_\ell[\sigma_\pm]^{\Tp_{\bar{A}}} \succeq 0, \\
      \chi_\ell[&\rho]_{\text{tr}} = 1, \quad  \chi_\ell [\sigma_-]_{\text{tr}} \le \Neg_0
\end{align}
\end{subequations}
where $\argmin_{\vecP}$ seeks for the argument minimizing the expression in \cref{Eq:Argmin}, $\vecf\reg^{\,(k)}$ is the regularized frequency obtained for the same iteration, and each $P(ab|xy)$ also appears as an optimization variable in the moment matrix $\chi_\ell[\rho]$. Then, the PBR used in the computation of $t_k$ can be evaluated by replacing $f\reg(ab|xy)$ and $P_\star(ab|xy)$ in \cref{Eq:PBR}, respectively, by $f\reg^{(k)}(ab|xy)$ and $P^{(k)}_\star(ab|xy)$. 

\begin{figure}[H]
    \centering
    \includegraphics[width=0.48\textwidth]{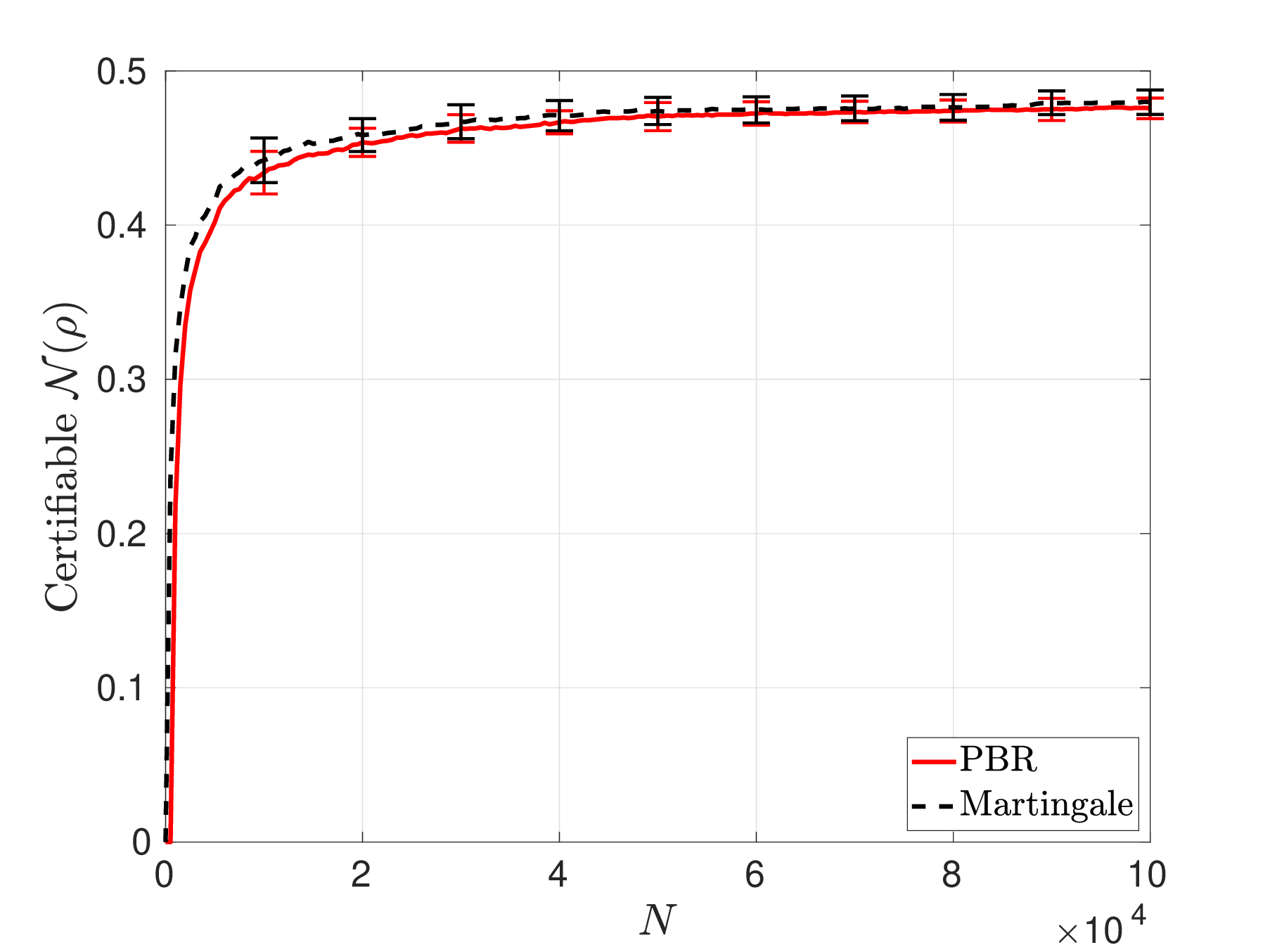}
    \caption{Negativity certifiable from the data observed in a Bell test generating $\PCHSH$, which arises by locally measuring the Bell state $\ket{\psi_\text{\tiny MES}}$ of \cref{Eq:MES2}. For the martingale-based protocol and any given $\Neg_0$ among $\bfN_0=\{0,0.01,\cdots,0.49\}$, we use \cref{Eq:Mart:CHSH:Neg} in \cref{Eq:mart2} to upper-bound $p\mart$ after every block of $\Nint=500$ trials, thereby generating $200\times 50$ upper bounds on $p\mart$ for a complete Bell test. For the PBR protocol and a given $\Neg_0$ from $\bfN_0$, we solve~\cref{Eq:PBR:Neg:PStar} by considering the same block size and the level-$3$ outer approximation of $\Q$ introduced in~\cite{MBL+13}. Then, we obtain $199\times 50$ upper bounds on $p\pbr$ from \cref{Eq:t1,Eq:tk,Eq:p-PBR}. To determine the lower bound on the underlying $\Neg(\rho)$ with the desired confidence of $\gamma\ge 99\%$, we look for the largest $\Neg_0$ in $\bfN_0$ such that $\H_{\Neg(\rho)\le\Neg_0}$ is rejected with a $p$-value bound being less than or equal to $0.01$. Each data point shown in the plot is an average over $30$ such lower bounds, and the error bar (standard deviation) gives an indication of the spread of the certifiable negativity. To avoid cluttering the plots, in each line, we show only a small number of markers.    
    \label{fig:CHSH_lv3_ga099}}
\end{figure}

In \cref{fig:CHSH_lv3_ga099}, we show the average amount of certifiable negativity from these two methods as a function of the number of trials $N$ employed. From the figure, it is clear that for certifying the underlying negativity using the data arising from $\PCHSH$, the performance of the two protocols are similar.
In fact, even though the martingale-based protocol appears to have a slight advantage over the PBR protocol for this certification task for small $N$'s, our computations of the asymptotic gain-rates $G\pbr$ and $G\mart$ show that they in fact agree (for all these values of $\Neg_0$ that we have considered), up to a numerical precision of $10^{-7}$. Also, in both cases, we see that with about $5\times10^3$ 
and $2\times10^4$ trials, 
we can already certify, respectively, more than $80\%$ and $90\%$ of the underlying negativity with a confidence $\gamma\ge0.99$. In \cref{Appdx.sec.CHSH}, we provide some additional plots showing how the $p$-value bound changes with $N$ for several values of $\Neg_0$.

\begin{figure}[H]
\centering
\includegraphics[width=0.48\textwidth,origin=c]{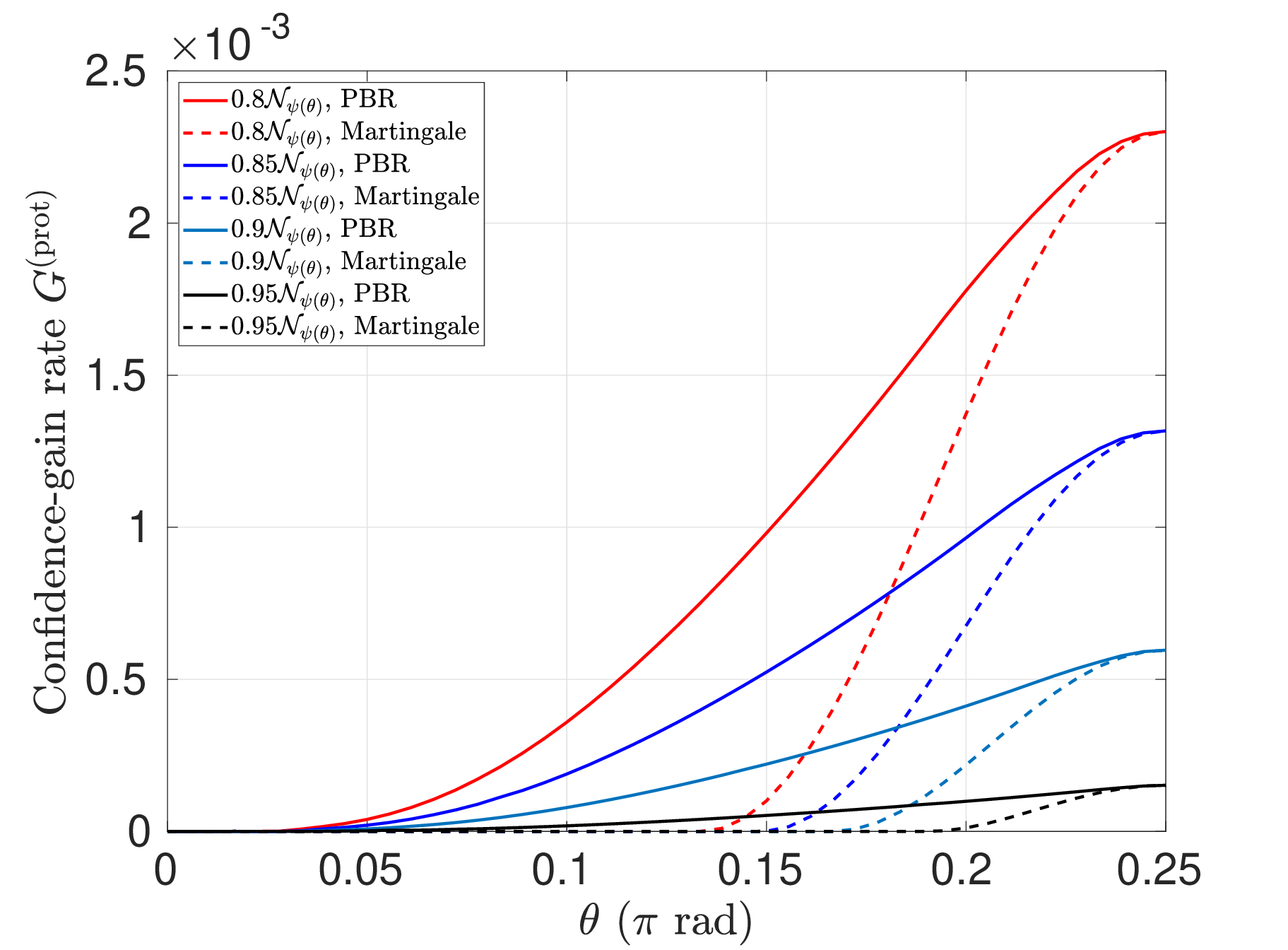} 
\caption{Asymptotic confidence-gain rate $G\prot$ based on the family of quantum correlations $\vecP_\theta$ derived from~\cref{Eq:Sefl-test:PES}, where $\theta = \frac{k\pi}{180}$ rad, $k=\{1,2,\cdots,45\}$, parametrizes the two-qubit entangled state $\ket{\psi(\theta)} = \cos{\theta} \ket{00} + \sin{\theta} \ket{11}$. Here, we again consider the composite hypothesis~\ref{H:Negativity}, with $\Neg_0=\{0.8\Neg_{\psi(\theta)}, 0.85\Neg_{\psi(\theta)}, 0.9\Neg_{\psi(\theta)}, 0.95\Neg_{\psi(\theta)}\}$ and $\Neg_{\psi(\theta)}$ being the negativity of $\psi(\theta)$, see \cref{Fig:NegBound}.
The gain rate for the martingale-based protocol is computed from \cref{Eq:Mart:G} using the CHSH Bell-like inequality of~\cref{Eq:Mart:CHSH:Neg} whereas that for the PBR protocol is evaluated from \cref{Eq:PBR:G} using the correlation derived from~\cref{Eq:Sefl-test:PES}.
}
\label{fig:CHSH_confidence-gainrates-rel}
\end{figure}

These results clearly suggest that the CHSH Bell function of \cref{Bellfunction_CHSH} is optimal for certifying the underlying negativity of $\ket{\psi_\text{\tiny MES}}$ using the martingale-based protocol. 
Indeed, a separate computation of \cref{Eq:KLD} and \cref{Eq:PBR} using $\PCHSH$ in place of $\vecf\reg$ show that, within a precision of $10^{-4}$, the optimized Bell-like inequality for $\Neg_0=0, 0.05,\cdots,1$ is equivalent to \cref{Eq:CHSH_Neg}.
How would things change if we perform a DI negativity certification using the data generated from the partially entangled state $\ket{\psi(\theta)}$, \cref{Eq:PsiTheta}? To this end, consider the quantum strategy of \cref{Eq:Sefl-test:PES}, whose resulting correlation $\vecP_\theta$ gives the maximal Bell CHSH violation for $\ket{\psi(\theta)}$, as well as the maximal violation of the tilted CHSH Bell inequality of~\cref{Ineq:TCHSH}. Then, instead of repeating the same analysis, we show in~\cref{fig:CHSH_confidence-gainrates-rel} the confidence-gain rates due to both protocols for certifying several {\em given fractions} of the underlying negativity. From the plots shown, it is evident that asymptotically, the martingale-based protocol employing the CHSH Bell function is far from optimal for certifying the underlying negativity of $\ket{\psi(\theta)}$. Indeed, the PBR protocol could identify some other Bell-like inequality that gives a much better confidence-gain rate, especially for the correlations arising from $\ket{\psi(\theta)}$ that is weakly entangled (small $\theta$). To a large extent, this can be understood by noting that the negativity lower bound of \cref{Eq:CHSH_Neg} due to its CHSH Bell violation is generally far from tight for these states, see~\cref{Fig:NegBound} in \cref{App:MISC}.

\subsubsection{Dimension certification via negativity certification}

As mentioned in \cref{Sec:NegDim}, the correlation $\PCHSH$ is insufficient to demonstrate any nontrivial dimension bound. Let us consider, instead, a correlation $\PCGLMP$ derived by local measuring the partially entangled two-qutrit state 
\begin{subequations}
\begin{equation}\label{Eq:Psi}
	\ket{\Psi} = \frac{1}{\sqrt{2+\zeta^2}}(\ket{00} + \zeta\ket{11} + \ket{22}),\quad \zeta = \frac{1}{2}(\sqrt{11}-\sqrt{3}),
\end{equation} 
with the local measurements
\begin{gather}
M^{(A)}_{a|x}=\ket{a}_{A,x}\bra{a}, \quad M^{(B)}_{b|y}=\ket{b}_{B,y}\bra{b}, \\
     \ket{a}_{A,x}  =  \sum^{2}_{j = 0} \frac{\omega^{j (\varphi^A_x+a)}}{\sqrt{3}} \ket{j},\quad 
     \ket{b}_{B,y}  = \sum^{2}_{j = 0} \frac{\omega^{j (\varphi^B_y-b)}}{\sqrt{3}}  \ket{j}, \nonumber
\end{gather}
\end{subequations}
where $\varphi^A_x =  \frac{x}{2}$, $\varphi^B_y =  (-1)^y\frac{1}{4}$, and $\{\ket{j}\}$ is the set of computational basis states. It is known~\cite{Acin2002,NPA2008} that this strategy gives the maximal CGLMP Bell-inequality violation of $\SCGLMP=1+\sqrt{\frac{11}{3}}\cong 2.91485$. Moreover, the negativity of $\ket{\Psi}$ can be easily evaluated to give $\cong 0.98358$.

Next, we use the numerically data simulated from $\PCGLMP$ to perform a hypothesis testing for Null Hypothesis~\ref{H:Negativity}, but now with $\Neg_0 \in \bfN_0=\{ 0.5, 0.51,\cdots,0.98 \}$. For the PBR protocol, the computation proceeds in exactly the same way as described above [see the paragraph containing~\cref{Eq:PBR:Neg:PStar}]. However, for the martingale-based protocol, since we do not have an explicit expression like that shown in~\cref{Eq:Mart:CHSH:Neg} for the CGLMP Bell expression, we compute an upper bound on $B_\H$ for each given value of $\Neg_0\in\bfN_0$ according to:
\begin{subequations}\label{Eq:BH:Neg}
\begin{align}
     \text{max} \,\, &\sum_{a,b,x,y} \beta^{ab}_{xy} P(ab|xy) \\
    \text{s.t. \ \ }  & \chi_\ell [\rho] = \chi_\ell[\sigma_+]-\chi_\ell[\sigma_-] , \quad \chi_\ell[\sigma_\pm]^{\Tp_{\bar{A}}} \succeq 0, 
    \label{Eq:SDP:BoundedNeg-Constraints}\\
     & \chi_\ell [\rho]\succeq 0,\quad \chi_\ell[\rho]_{\text{tr}} = 1,\quad  \chi_\ell[\sigma_-]_{\text{tr}}\le \Neg_0,   
\end{align}
\end{subequations}
where the CGLMP Bell coefficients $\beta^{ab}_{xy}$ are defined in \cref{Ineq:CGLMP}. Meanwhile, since $P_{xy}=\frac{1}{4}$ and $\beta^{ab}_{xy}\in\{-1,0,1\}$, we again have $\mathfrak{b}_\pm=\pm4$ for $\ICGLMP$.

\begin{figure}[H]
\centering
\includegraphics[width=0.48\textwidth]{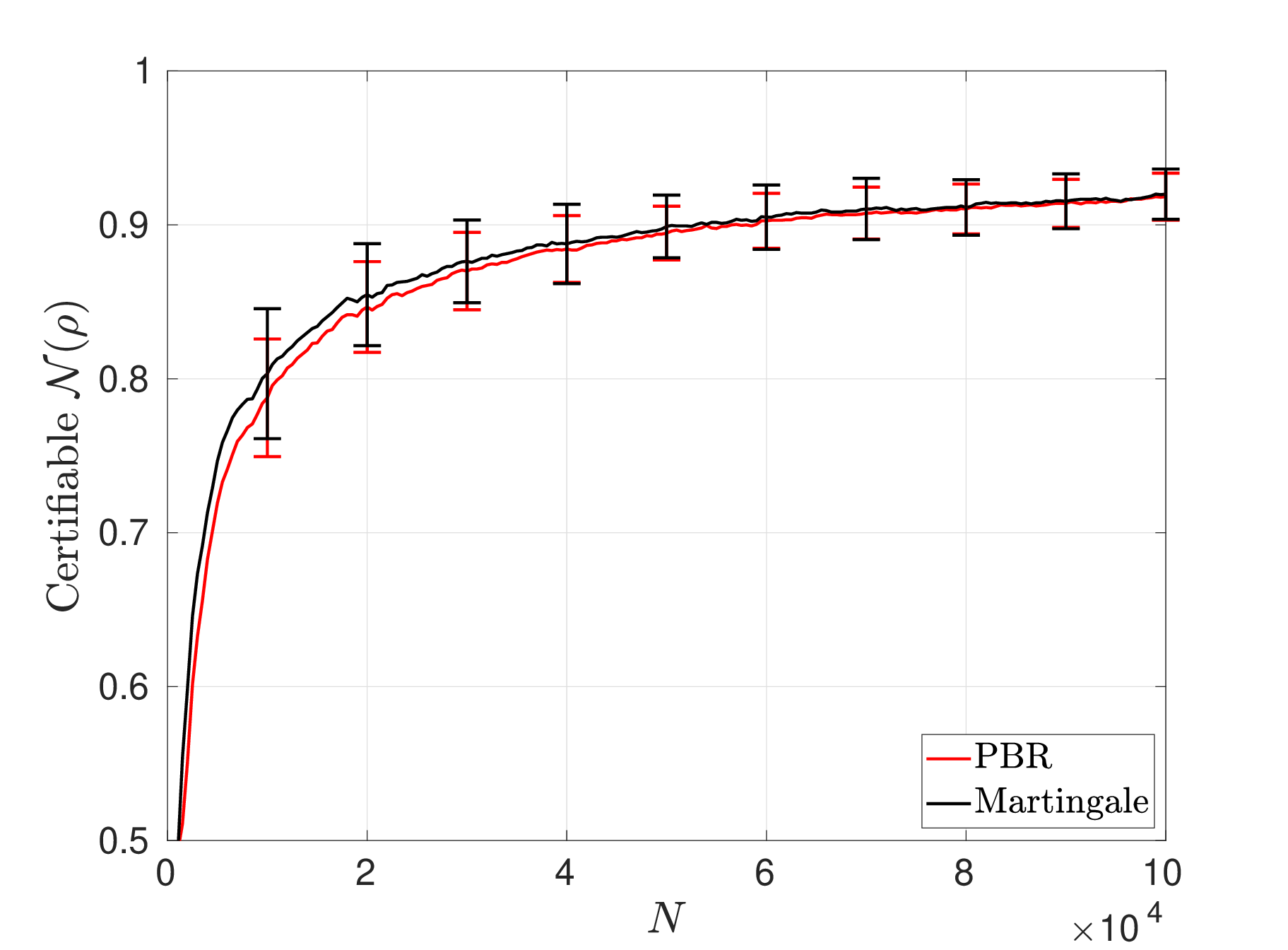} 
\caption{Negativity certifiable from the data observed in a Bell test generating $\PCGLMP$, which arises by locally measuring the partially entangled state of \cref{Eq:Psi}. For details on how the plot is generated, see the caption of \cref{fig:CHSH_lv3_ga099}, but bearing in mind that for the martingale-based protocol, we now use in \cref{Eq:mart2} the $B_\H$ determined from \cref{Eq:BH:Neg}.\label{fig:CGLMP_lv14_ga099_100k}}
\end{figure}

\begin{figure}[H]
\centering
\includegraphics[width=0.48\textwidth]{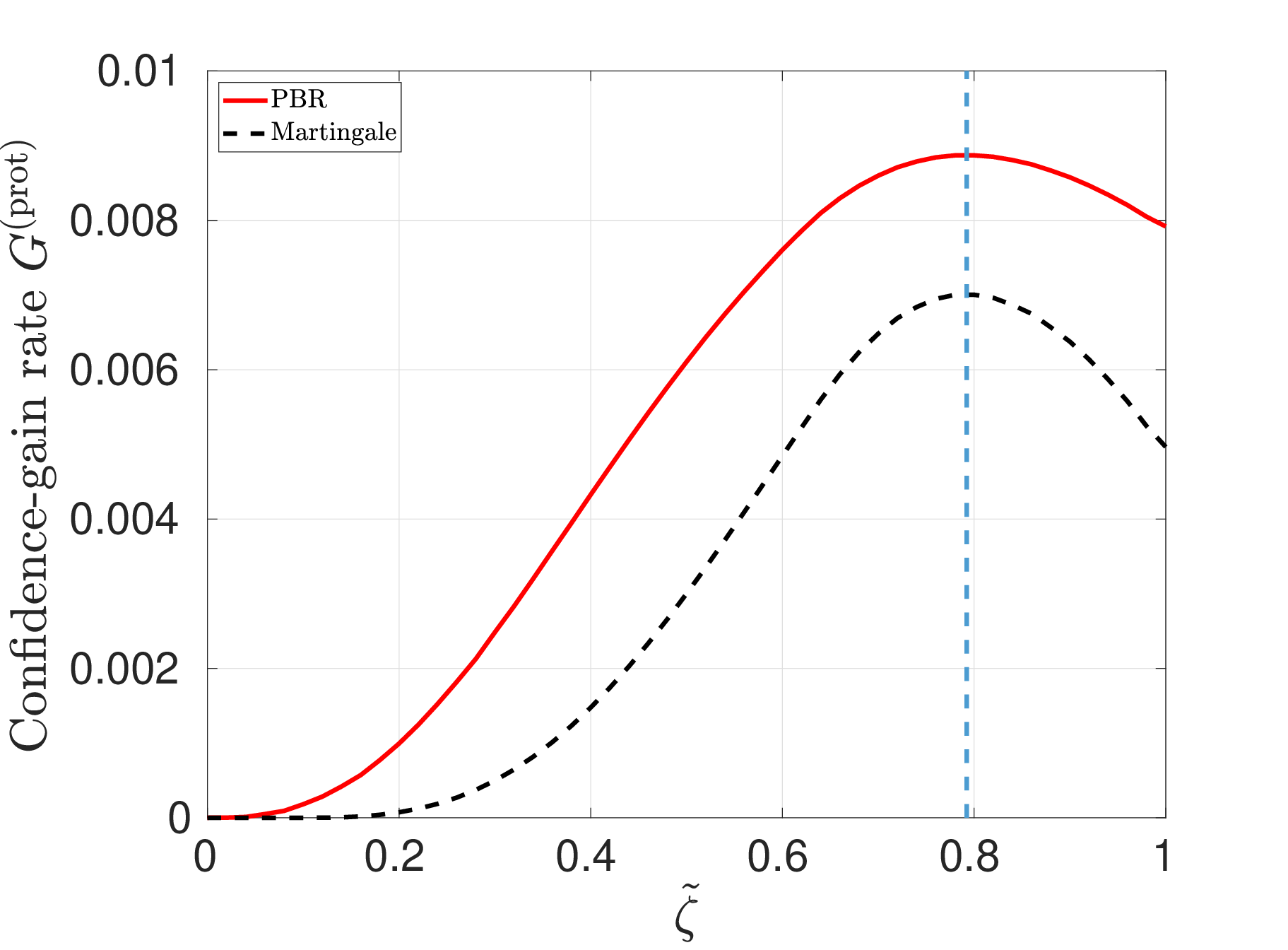} 
\caption{
Asymptotic confidence-gain rate $G\prot$ based on the quantum correlation $\vecP_{\tilde{\zeta}}$ derived from maximizing the CGLMP Bell-inequality-violation of $\ket{\Psi(\tilde{\zeta})}$, \cref{Eq:PsiZeta}. Here, we consider $\tilde{\zeta}=\{0, 0.02,...,1\}$ and the composite hypothesis~\ref{H:Negativity}, with $\Neg_0=0.5$. 
The gain rate for the martingale-based protocol is computed from \cref{Eq:Mart:G} using the CGLMP Bell function and the bound $B_\H$ determined in \cref{Eq:BH:Neg} whereas that for the PBR protocol is evaluated from \cref{Eq:PBR:G} and setting $\vecP_\Q$ to $\vecP_{\tilde{\zeta}}$.  The dashed blue line corresponds to the value $\tilde{\zeta}=\zeta$, cf.~\cref{Eq:Psi}. }
\label{fig:CGLMP02 compare gain rate L2}
\end{figure}

From \cref{fig:CGLMP_lv14_ga099_100k}, we see that with about $6\times10^4$ trials
 we can already certify a negativity lower bound of $0.9$.  On the other hand, if we want to certify that we need at least a {\em two-qutrit state} to produce the observed data (arising from $\PCGLMP$), it suffices to certify that the underlying negativity is {\em strictly} larger than $0.50$, which happens already with approximately 1500 trials. 
 Could other two-qutrit states provide a more favorable correlation in this regard? To gain insight into the prolem, we consider the following one-parameter family of two-qutrit states\begin{equation}\label{Eq:PsiZeta}
	\ket{\Psi(\tilde{\zeta})} = \frac{1}{\sqrt{2+\tilde{\zeta}^2}}(\ket{00} + \tilde{\zeta} \ket{11} + \ket{22})
\end{equation}
and numerically maximize their CGLMP Bell-inequality violation using the heuristic algorithm given in~\cite{LD07}. We denote the corresponding correlation by~$\vecP_{\tilde{\zeta}}$, compute the corresponding asymptotic confidence-gain rate for both protocols, and plot the results in \cref{fig:CGLMP02 compare gain rate L2}.

Interestingly, even though \cref{fig:CGLMP_lv14_ga099_100k} suggests that the CGLMP Bell function is very effective in providing a good $p$-value bound against hypothesis~\ref{H:Negativity}, ~\cref{fig:CGLMP02 compare gain rate L2} clearly show that, asymptotically, it is not optimal. The results shown in~\cref{fig:CGLMP02 compare gain rate L2} further suggest that among the family of two-qutrit states given in~\cref{Eq:PsiZeta}, the qutrit signature of $\ket{\Psi(\zeta)}$, cf. \cref{Eq:Psi}, could even be the most prominent, when it comes to its DI certification using these hypothesis-testing techniques.

\subsection{Entanglement depth certification}
\label{Sec:Results:ED}

Next, we consider the tripartite correlation $\vecP_\text{GHZ}$ that results from  locally measuring the $\pm1$-eigenvalue observables
\begin{subequations}\label{Eq:GHZ:Strategies}
\begin{equation}
A_0=B_0=C_0=\sigma_y, \quad A_1=B_1=C_1=-\sigma_x
\end{equation}
on the Greenberger-Horne-Zeilinger (GHZ) state~\cite{GHZ,Mermin90AJP}:
\begin{equation}
	\ket{\text{GHZ}} = \frac{1}{\sqrt{2}}(\ket{000}+\ket{111}).
\end{equation}
\end{subequations}
It is easy to verify that $\vecP_\text{GHZ}$ leads to a violation of the Mermin Bell inequality, ~\cref{Ineq:Mermin,Eq:Mermin:Bounds}, giving the algebraic maximum of $\SMermin=4$. For our simulations, we assume a uniform distribution $P_{xyz}=\frac{1}{4}$ over all measurement settings $x,y,z\in\{0,1\}$ that satisfy $\text{mod}(x+y+z,2)=1$. Then, we test the data against the following composite hypotheses.
\begin{hypothesis}\label{H:Sep}
	$\H_\text{Sep}$: In every experimental trial, the underlying state is separable (having an entanglement depth of $1$). 
\end{hypothesis}
\begin{hypothesis}\label{H:BiSep}
	$\H_\text{2-prod}$: In every experimental trial, the underlying state is 2-producible, i.e., having an entanglement depth of $2$ or less. 
\end{hypothesis}

For the martingale-based method, we use \cref{Eq:mart2} with the Mermin Bell expression of \cref{Ineq:Mermin} and the bounds given in \cref{Eq:Mermin:Bounds}, i.e., $B_\H = 2$ for hypothesis~\ref{H:Sep} and $B_\H = 2\sqrt{2}$ for hypothesis~\ref{H:BiSep}. Since $\beta^{abc}_{xyz}\in\{-1,0,1\}$, we again have $\mathfrak{b}_\pm=\pm4$. Note that separable states can only generate Bell-local correlations~\cite{BCP+14}, cf.~\cref{Eq:BellLocal}. Thus, for the PBR protocol with hypothesis~\ref{H:Sep}, 
the optimizing distribution $P^{(k)}_\star(abc|xyz)$ for the $k$-iteration can be obtained by solving [cf. \cref{Eq:KLD}]
\begin{subequations}\label{Eq:PBR:Local:PStar}
\begin{align}
      \argmin_{\vecP} \,\,&-\!\!\!\!\!\!\sum_{a,b,c,x,y,z}\!\!\!\!\! P_{xyz} f\reg^{(k)}(abc|xyz) \log P(abc|xyz),\label{Eq:Argmin3}\\
    \text{s.t.}\,\,  \vecP &= \sum_i q_i \vec{D}_i,\quad q_i\ge 0, \quad \sum_i q_i=1
\end{align}
\end{subequations}
where $\vec{D}_i$ is the $i$-th (local deterministic) extreme points of the set of tripartite Bell-local distributions. 

On the other hand, notice that 2-producibility~\cite{Guhne:NJP:2005} is equivalent to biseparability~\cite{HHHH09} in the tripartite scenario. Hence, for hypothesis~\ref{H:BiSep}, we obtain the corresponding optimizing distribution by solving
\begin{subequations}\label{Eq:PBR:Bisep:PStar}
\begin{align}
      \argmin_{\vecP} \,\,&-\!\!\!\!\!\!\sum_{a,b,c,x,y,z}\!\!\!\!\! P_{xyz} f\reg^{(k)}(abc|xyz) \log P(abc|xyz),\label{Eq:Argmin3b}\\
    \text{s.t.} \; & \chi_\ell [\rho] = \chi_\ell [\rho_1] + \chi_\ell [\rho_2] + \chi_\ell [\rho_3], \, \chi_\ell [\rho] \succeq 0, \\
    &  \chi_\ell [\rho]_{\text{tr}} = 1,\quad \chi_\ell [\rho_i] \succeq 0, \quad \forall\,\, i \in \{1,2,3\}, \\
    & \chi_\ell [\rho_1]^{\Tp_{\bar{A}}} \succeq 0, \, \chi_\ell [\rho_2]^{\Tp_{\bar{B}}} \succeq 0, \, \chi_\ell [\rho_3]^{\Tp_{\bar{C}}} \succeq 0,
    \label{bisepSDP}
\end{align}
\end{subequations}
where $\rho_i$, with $i = \{ 1,2,3 \}$ are meant to represent, respectively, the constituent of $\rho$ that is separable with respect to the $A|BC$, $B|AC$, and $C|AB$ bipartitions. In evaluating \cref{Eq:PBR:Bisep:PStar}, we use level $\ell=1$ of the hierarchy introduced in \cite{MBL+13}. For both hypotheses, we then evaluate 
\begin{equation}
	R(a,b,c,x,y,z)=\frac{f\reg^{(k)}(abc|xyz)}{P_\star(abc|xyz)}
\end{equation}
for the computation of the test statistic $t_k$. 

For $\vecP_\text{GHZ}$ and hypothesis~\ref{H:Sep}, the confidence-gain rate $G\pbr$ is already known (see Table I of~\cite{vDGG05}) to be approximately $0.415037$; our computation reproduces this and further shows that for hypothesis~\ref{H:BiSep}, this is approximately $0.228446$. Moreover, to six decimal places, $G\pbr$ and $G\mart$ agree for both hypotheses. What about finite data? Based on the average results from $30$ simulations, we find that the $p$-value bounds, or more precisely, $\Pf=-\log_2p\prot$can be very well fitted into the following straight lines:\footnote{In all these fits, the coefficient of determination $R^2$ is $1$ even if we keep up to 7 significant digits.}
\begin{gather}
    \Pf\pbr_{\H_\text{Sep}} = 0.414958\, N -204.978,\,\, N\in[10^3, 10^5], \\
    \Pf\mart_{\H_\text{Sep}} = 0.415037\, N,\quad N\in[0, 10^5] 
\end{gather}
for the separable hypothesis $\H_\text{Sep}$ of \ref{H:Sep}, and 
\begin{gather}
    \Pf\pbr_{\H_\text{2-prod.}} = 0.22838\,N -115.22,\,\, N\in[10^3, 10^5], \\
    \Pf\mart_{\H_\text{2-prod.}} = 0.228447\, N,\quad N\in[0, 10^5] 
\end{gather}
for the 2-producible hypothesis $\H_\text{2-prod.}$ of \ref{H:BiSep}. Consequently, based on this interpolation, even if we only run the Bell test using the strategy of \cref{Eq:GHZ:Strategies} for 100 trials, there is already sufficient data to certify  genuine tripartite-entanglement with a confidence of at least $1-10^{-6}$.

\subsection{Fidelity certification}
\label{Sec:Res:Fidelity}

Our last examples concern the DI certification of a lower bound on the fidelity of the swapped state $\rhoS$ with respect the target state $\ket{\psi(\theta)}$ of \cref{Eq:PsiTheta}. To this end, we use the same set of data generated for the analysis in \cref{Subsec. CHSH} and 
consider the following null hypothesis.
\begin{hypothesis}\label{H:Fid}
	$\H_{\F_\theta(\rhoS)\le \F_0}$: In every experimental trial, the swapped state $\rhoS$ extractable from the underlying state $\rho$ has a $\ket{\psi(\theta)}$-fidelity upper bounded by $\F_0$, i.e., 
	\begin{equation}
		\F_\theta(\rhoS):= \bra{\psi(\theta)} \rhoS\ket{\psi(\theta)} \le \F_0.
	\end{equation}
\end{hypothesis}

Then, for any given $\theta$ and $\F_0\ge \cos\theta$, to apply the PBR protocol, we solve the optimizing distribution $P^{(k)}_\star(ab|xy)$ for the $k$-iteration [cf. \cref{Eq:KLD}] via:
\begin{subequations}\label{Eq:PBR:Fid:PStar}
\begin{align}
      \argmin_{\vecP} \,\,&-\sum_{a,b,x,y} P_{xy} f\reg^{(k)}(ab|xy) \log P(ab|xy),\label{Eq:Argmin:F}\\
    \text{s.t. \ \ }  \chi_\ell [&\rho] \succeq 0, \quad \chi_\ell[\rho]_{\text{tr}} = 1,\quad \F_\theta(\rhoS) \le \F_0,\label{Eq:Constraints:Fid}
\end{align}
\end{subequations}
where the left-hand side of the last inequality in~\cref{Eq:Constraints:Fid} consists of some specific linear combination of entries of $\chi_\ell[\rho]$, see~\cite{Bancal15} for details. Then, as with negativity certification, we can evaluate the PBR used in the computation of $t_k$ by replacing $f\reg(ab|xy)$ and $P_\star(ab|xy)$, respectively, by $f\reg^{(k)}(ab|xy)$ and $P^{(k)}_\star(ab|xy)$ in \cref{Eq:PBR}. As for the martingale-based protocol, we first solve 
\begin{subequations}\label{Eq:BH:fid}
\begin{align}
     \text{max} \,\, &\sum_{a,b,x,y} \beta^{ab}_{xy} P(ab|xy) \\
    \text{s.t. \ \ }  \chi_\ell [&\rho] \succeq 0, \quad \chi_\ell[\rho]_{\text{tr}} = 1,\quad \F_\theta(\rhoS) \le \F_0
\end{align}
\end{subequations}
to determine $B_{\H}$ for hypothesis~\ref{H:Fid} and then apply \cref{Eq:mart2} to determine the corresponding $p$-value upper bound.

\begin{figure}[H]
\centering
\includegraphics[width=0.45\textwidth]{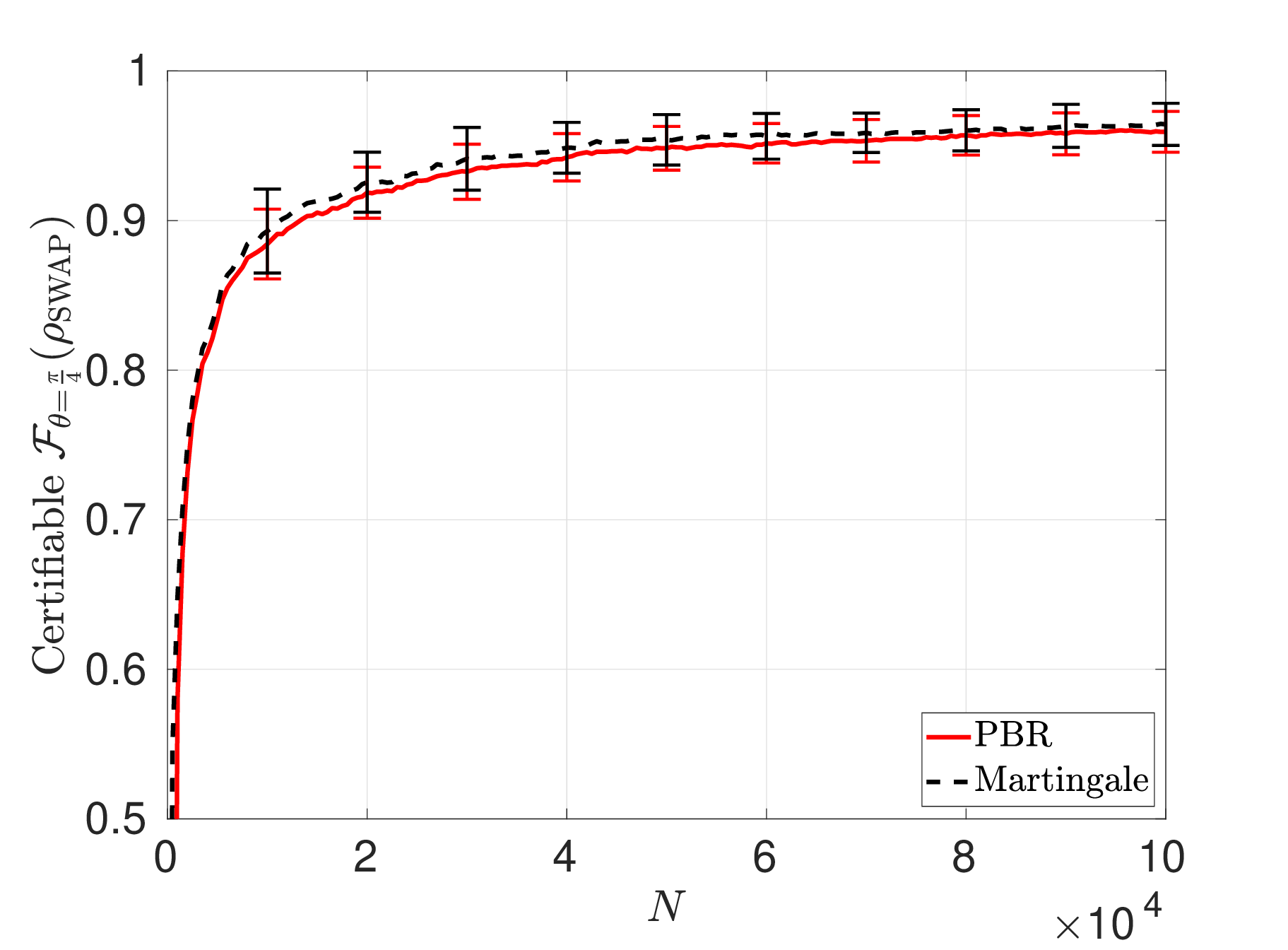} 
\caption{
Certifiable fidelity $\F_{\theta=\frac{\pi}{4}}(\rhoS)$  from the data observed in a Bell test generating $\PCHSH$, which arises by locally measuring the Bell state $\ket{\psi_\text{\tiny MES}}$ of \cref{Eq:MES2}. For the martingale-based protocol and any given $\F_0$ among $\bmF_0=\{0.5,0.51,\cdots,0.99\}$, we use the $B_\H$ determined from~\cref{Eq:BH:fid} in \cref{Eq:mart2} to upper-bound $p\mart$ after every block of $\Nint=500$ trials, thereby generating $200\times 50$ upper bounds on $p\mart$ for a complete Bell test. For the PBR protocol and a given $\F_0$ from $\bmF_0$, we solve~\cref{Eq:PBR:Fid:PStar} by considering the same block size and the level-$2$ outer approximation of $\Q$ introduced in~\cite{MBL+13}. Then, we obtain $199\times 50$ upper bounds on $p\pbr$ from \cref{Eq:t1,Eq:tk,Eq:p-PBR}. To determine the lower bound on the underlying $\F_{\theta=\frac{\pi}{4}}(\rhoS)$ with the desired confidence of $\gamma\ge 99\%$, we look for the largest $\F_0$ in $\bmF_0$ such that $\H_{\F_\theta(\rhoS)\le \F_0}$ is rejected with a $p$-value upper bound less than or equal to $0.01$. 
}
\label{fig:CHSH_extractability_ga099}
\end{figure}

Let us start with the self-testing of a Bell state, corresponding to $\theta=\frac{\pi}{4}$ in \cref{Eq:PsiTheta}. In this case, we use the CHSH Bell function specified in~\cref{Bellfunction_CHSH}  and consider $\F_0\in\bmF_0 = \{0.5,0.51,\cdots,0.99\}$. For both protocols, by systematically evaluating the $p$-value bounds from the data for each of these $\F_0$'s, we determine a lower bound on $\F_{\theta=\frac{\pi}{4}}(\rhoS)$ with the desired confidence of at least $99\%$. The results obtained from both hypothesis-testing protocols are shown in~\cref{fig:CHSH_extractability_ga099}.

\begin{figure}[H]
\centering
\includegraphics[width=0.48\textwidth]{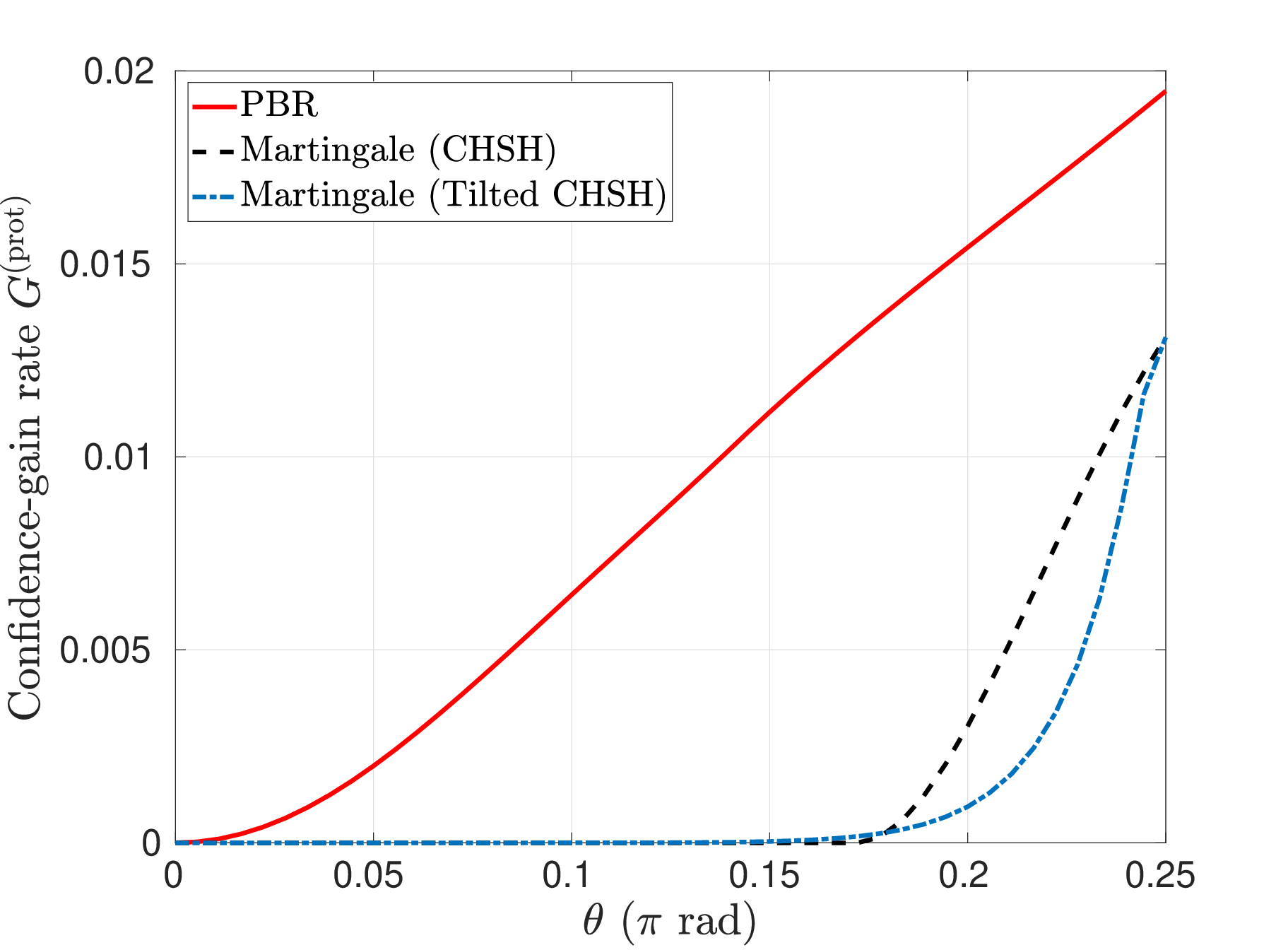} 
\caption{
Asymptotic confidence-gain rate $G\prot$ based on the family of quantum correlations $\vecP_\theta$ derived from~\cref{Eq:Sefl-test:PES}, where $\theta = \frac{k\pi}{180}$ rad, $k=\{1,2,\cdots,45\}$. Here, we consider Null Hypothesis~\ref{H:Fid}, with $\F_\theta(\rhoS)=\cos^2\theta$, the trivial fidelity achievable without shared entanglement.
The gain rate for the martingale-based protocol is computed from \cref{Eq:Mart:G} using the CHSH Bell function of~\cref{Bellfunction_CHSH} (dashed line, black) and the tilted CHSH Bell function of~\cref{Ineq:TCHSH} (dashed-dotted line, blue) whereas that for the PBR protocol is evaluated from \cref{Eq:PBR:G}. 
}
\label{fig:CHSH_extractability_confidencegain}
\end{figure}

Interestingly, our results show that the martingale-based protocol with the CHSH Bell function of~\cref{Bellfunction_CHSH} again performs very well for the self-testing of a Bell state with finite statistics, even though our computation of the corresponding asymptotic confidence-gain rate for $\F_0=0.5$ clearly shows that it is suboptimal even for the Bell state, see~\cref{fig:CHSH_extractability_confidencegain}. What about other partially entangled states? To answer this question, we evaluate the confidence-gain rate derived from both protocols for\footnote{For $\ket{\psi{(\theta})}$ defined in \cref{Eq:PsiTheta}, a fidelity of $\cos^2\theta$ is always achievable even if Alice and Bob does not share any entanglement; they merely have to prepare $\ket{00}$ using local operations and classical communication before the Bell test.}  $\F_0=\cos^2\theta$ with $\theta = \{0^\circ, 1^\circ, 2^\circ, ..., 45^\circ\}$. This time around, for the martingale-based protocol, we switch to the Bell function of the tilted CHSH Bell inequality of \cref{Ineq:TCHSH}, which is known to facilitate the self-testing of all entangled $\ket{\psi(\theta)}$.  The corresponding results are shown in~\cref{fig:CHSH_extractability_confidencegain}.

\subsection{Properties certification via Bell-value certification}\label{Sec:PropertiesViaCHSH}

The advantage of a fidelity certification based on the SWAP method~\cite{YVB+14,Bancal15} is that the technique is applicable to a general Bell scenario. However, in the simplest CHSH Bell scenario, it is known that a much tighter lower bound on the Bell-state fidelity can be obtained by considering a more general extraction map. Specifically, Kaniewski showed in~\cite{Kaniewski2016} that
\begin{equation}\label{Eq:TighterFidelity}
\begin{aligned}
	\max_{\Lambda_A,\Lambda_B} \min_{\rho_{AB}} \F[(\Lambda_A \otimes \Lambda_B)(\rho_{AB}),\ket{\psi_\text{MES}}\!\!&\bra{\psi_\text{MES}}])\ge \\
	&\frac{1}{2}+\frac{1}{2}\frac{\SCHSH-\beta^*}{2\sqrt{2}-\beta^*},
\end{aligned}
\end{equation}
where $\Lambda_A$, $\Lambda_B$ are local extraction maps acting, respectively, on Alice's and Bob's subsystem, while  $\beta^*:=\frac{16+14\sqrt{2}}{17}\approx 2.1058$ is the threshold CHSH value for which the fidelity bound becomes trivial.

To take advantage of \cref{Eq:TighterFidelity}, we can first perform a hypothesis testing based on the following null hypothesis.
\begin{hypothesis}\label{H:CHSH}
	$\H_{\SCHSH(\rho)\le \S_0}$: In every experimental trial, the underlying state and measurements give a CHSH value $\SCHSH$ less than or equal to $\S_0$.
\end{hypothesis}
Specifically, using the same set of data  generated for the analysis in \cref{Subsec. CHSH} and \cref{Sec:Res:Fidelity}, we perform composite hypothesis testing for Null Hypothesis \ref{H:CHSH} with 
$\S_0\in\{2, 2+\Delta\S, 2+2\Delta\S, \cdots, 2\sqrt{2}-\Delta S\}$, where $\Delta\S = \frac{2(\sqrt{2}-1)}{50}$.

\begin{figure}[H]
\centering
\includegraphics[width=0.45\textwidth]{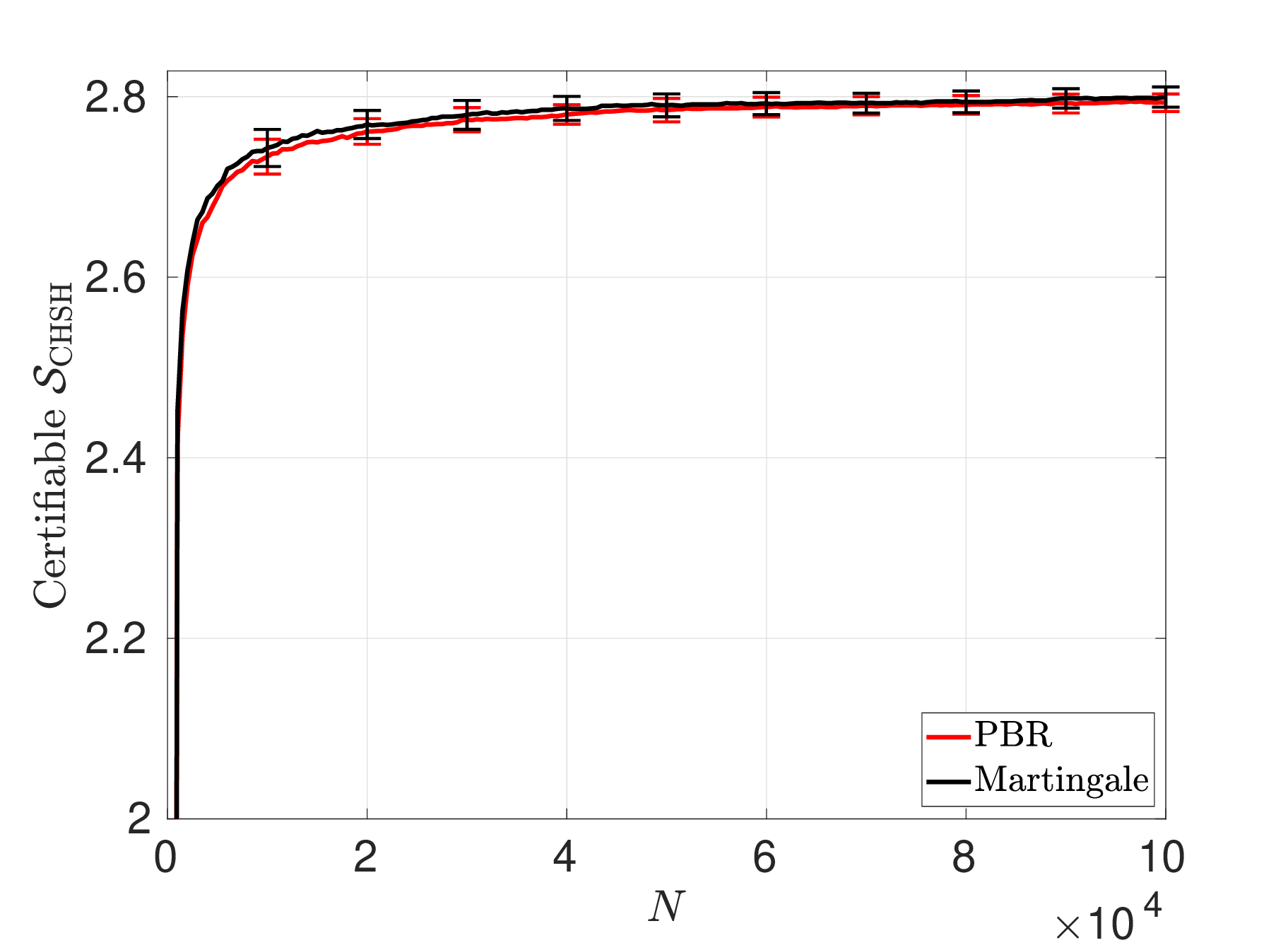} 
\caption{
Certifiable Bell-CHSH violation $\SCHSH$  from the data observed in a Bell test generating $\PCHSH$, which arises by locally measuring the Bell state $\ket{\psi_\text{\tiny MES}}$ of \cref{Eq:MES2}. For the martingale-based protocol and any given $\S_0$ among $\bm{S}_0=\{2+k\Delta\S\}_{k=0}^{49}$, we use \cref{Eq:Constraints:S} in \cref{Eq:mart2} to upper-bound $p\mart$ after every block of $\Nint=500$ trials, thereby generating $200\times 50$ upper bounds on $p\mart$ for a complete Bell test. For the PBR protocol and a given $\S_0$ from $\bm{S}_0$, we solve~\cref{Eq:PBR:S:PStar} by considering the same block size. Then, we obtain $199\times 50$ upper bounds on $p\pbr$ from \cref{Eq:t1,Eq:tk,Eq:p-PBR}. To determine the lower bound on the underlying $\SCHSH$ with the desired confidence of $\gamma\ge 99\%$, we look for the largest $\S_0$ in $\bm{S}_0$ such that $\H_{\SCHSH\le\S_0}$ is rejected with a $p$-value bound being less than or equal to $0.01$. A separate calculation shows that if we impose \cref{Eq:SDP:Quantum} in addition to \cref{Eq:Constraints:S}, one may find a visually indistinguishable difference ($<5\times10^{-4}$) in the certifiable $\SCHSH$. 
}
\label{fig:CHSH}
\end{figure}

In particular, for the martingale-based protocol, we can simply use \cref{Eq:mart2} with $\mathfrak{b}_\pm=\pm4$ and $B_\H=\S_0$. Could one also employ the PBR protocol, which does not usually presuppose any Bell-like inequality, for the current hypothesis testing? This is indeed possible. To this end, one may solve the optimizing distribution $P^{(k)}_\star(ab|xy)$ for the $k$-iteration [cf. \cref{Eq:KLD}] of the PBR protocol via:
\begin{subequations}\label{Eq:PBR:S:PStar}
\begin{align}
      \argmin_{\vecP} \,\,&-\sum_{a,b,x,y} P_{xy} f\reg^{(k)}(ab|xy) \log P(ab|xy),\label{Eq:Argmin:S}\\
    \text{s.t. \ \ }  &\sum_{a,b,x,y} (-1)^{xy+a+b} P(ab|xy) \le \S_0,\label{Eq:Constraints:S}
\end{align}
\end{subequations}
with or without imposing the SDP constraints of \cref{Eq:SDP:Quantum}. The results obtained from these tests are shown in~\cref{fig:CHSH}.

\begin{figure}[H]
\centering
\includegraphics[width=0.45\textwidth]{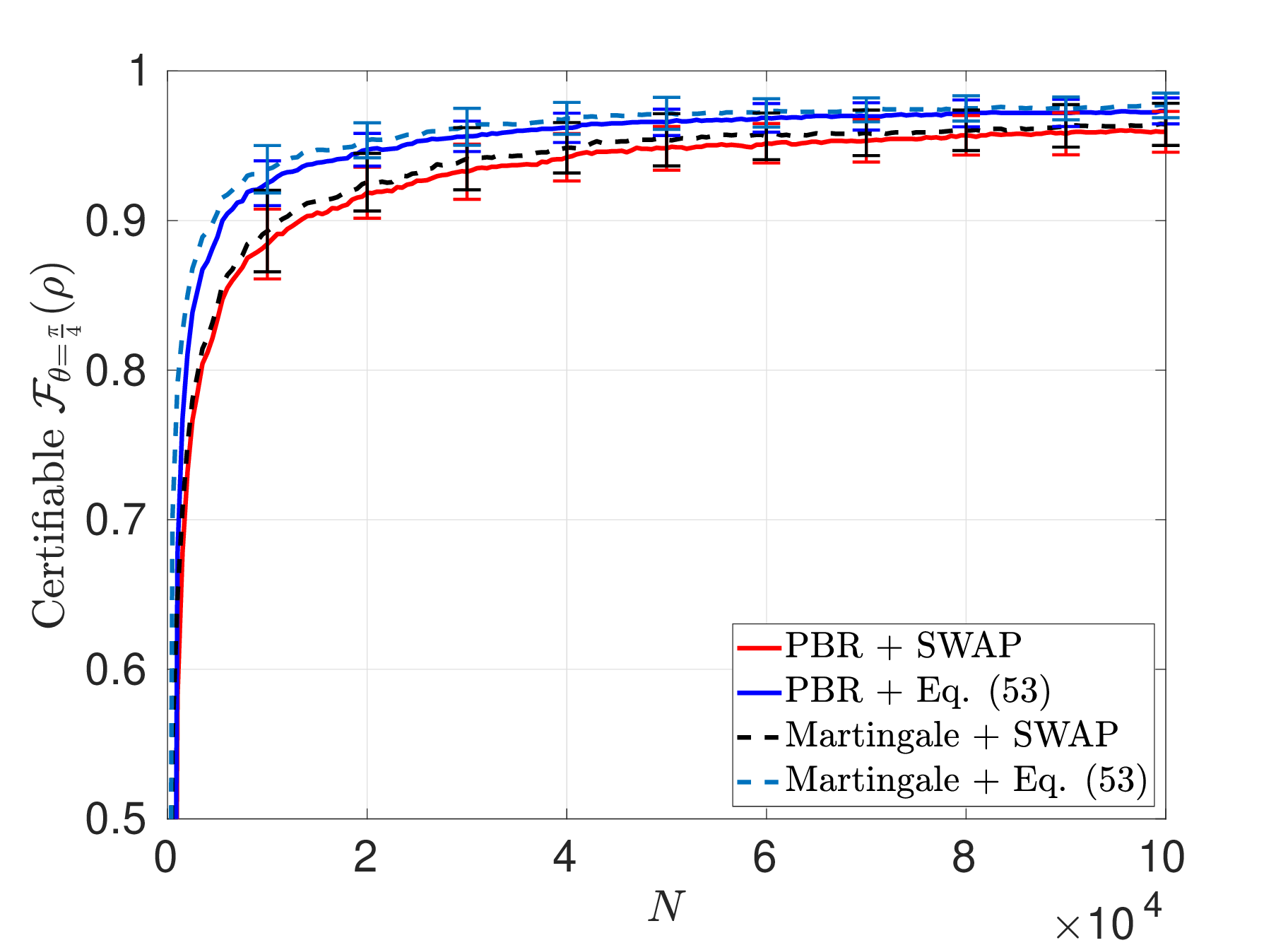} 
\caption{
Comparison of the Bell-state fidelity certifiable via the SWAP-based approach presented in~\cref{Sec:Res:Fidelity} and that via \cref{Eq:PBR:S:PStar} and $\SCHSH$-value certification.
}
\label{fig:fidelity-all}
\end{figure}

Using each lower bound on $\SCHSH$ certified from the data, \cref{Eq:TighterFidelity} immediately translates to a lower bound on the Bell-state fidelity with the desired confidence. For a direct comparison with the efficacy of the SWAP-based approach adopted in~\cref{Sec:Res:Fidelity}, we plot in~\cref{fig:fidelity-all} the Bell-state fidelity certifiable using the two approaches. As expected, the tighter Bell-state fidelity lower bound provided by \cref{Eq:TighterFidelity} also facilitates a considerably tighter lower bound when one has access to only a finite amount of data. 

It is also worth noting that in computing these PBR bounds, the computation may be further simplified by regularizing the relative frequency $\vecf\reg$ using {\em only} the nonsignaling constraint of \cref{Eq:NS}, instead of the quantum approximation via \cref{Eq:SDP:Quantum}. For the lower bounds on $\SCHSH$ presented in \cref{fig:CHSH}, this further simplification was found to give, unsurprisingly, a worse lower bound, but with a deviation bounded by $8\times10^{-3}$. Of course, the lower bounds on $\SCHSH$ can also be used to bound other desired properties. For example, \cref{fig:CHSH_lv3_ga099} can equivalently be obtained by combining~\cref{Eq:CHSH_Neg} with the results shown in~\cref{fig:CHSH}.

\section{Discussion}\label{Sec:Discussion}

Tomography and witnesses are two commonly employed toolkits for certifying the desirable properties of quantum devices~\cite{Eisert:2020aa}. In recent years, the device-independent (DI) paradigm has offered an appealing alternative to these conventional means as it involves only a minimal set of assumptions. Nonetheless, many DI certification schemes, e.g.,~\cite{BGLP11,MBL+13,TMG15,LRB+15,CKZ+21,CBLC16,CBLC18,Bancal:PRL:2018,Renou:PRL:2018,QBW+19,WBSS20,Sekatski2018,SNI+23}, implicitly assumes that the underlying quantum correlation $\vecP_\Q$ (or the actual Bell-inequality violation due to $\vecP_\Q$) is known. In practice, this is unrealistic for two reasons: (1) we always have access to only a finite amount of experimental data, and (2) actual experimental trials are typically {\em not} independent and identically distributed ({\em i.i.d.}).

To this end, very specialized tools have been developed for the task of randomness generation, quantum key distributions, and the self-testing~\cite{Bancal15,TWE+17,BRS+21} of quantum states. Among them, the possibility of using hypothesis testing (based on the PBR protocol~\cite{ZGK11}) for self-testing with finite data was first discussed in~\cite{Bancal15} (see also~\cite{TWE+17} for a different approach). Meanwhile, it is long known~\cite{Gill2003,Gill2003b,ZGK13} that hypothesis testing in a Bell test can also be carried out using a martingale-based protocol. Here, we demonstrate the viability and versatility of such hypothesis-testing-based approaches for the general problem of DI certification. 

Central to our finding is the observation that many desirable quantum properties $\mathcal{P}$ that one wishes to certify can be characterized by (the {\em complement} of) some convex set $\mathcal{C}_\P$ in the space of correlation vectors $\{\vecP\}$. In other words, if a given $\vecP_\Q$ lies outside $\mathcal{C}_\P$, a Bell-like inequality can be provided to witness this fact. This separating hyperplane then provides the basis for our martingale-based protocol for DI certification. On the other hand, if $\mathcal{C}_\P$ itself admits a semidefinite programming characterization like the kind proposed in~\cite{NPA,NPA2008,DLTW08,MBL+13,CBLC16}, then the problem of minimizing the statistical distance to $\mathcal{C}_\P$ can be cast as a conic program, which can be readily solved using existing solvers, such as $\mathsf{MOSEK}$~\cite{Mosek_exponential_cone}. In turn, the PBR protocol provides an {\em optimized Bell-like inequality} that facilitates the corresponding hypothesis testing. 

In this paper, we explain in detail how the two aforementioned hypothesis-testing protocols can be adapted for DI certification of desirable properties. Specifically, we illustrate how we can use them to perform DI certification of the underlying negativity~\cite{Vidal02}, local Hilbert space dimension~\cite{BPA+08}, entanglement depth~\cite{Guhne:NJP:2005,Sorensen2001May}, and fidelity to some target two-qubit entangled pure state $\ket{\psi(\theta)}$. In each of these examples, we further demonstrate how the certifiable property (with a confidence of $99\%$) varies with the number of experimental trials involved, see \cref{fig:CHSH_lv3_ga099}, \cref{fig:CGLMP_lv14_ga099_100k}, \cref{fig:CHSH_extractability_ga099}. Even though we have focused on certifying desirable properties of quantum states, as explained above, the protocols can also be applied to certify desirable properties of the measurement devices, such as their measurement incompatibility~\cite{CBLC16,CBLC18,QBW+19,CMBC21}, or their similarity to some target measurements~\cite{Bancal15} or instruments~\cite{WBSS20}, etc. Note, however, that the usefulness of our protocols relies on the possibility of certifying the desired property from a Bell-inequality-violating correlation. To this end, we remind that determining the complete list of quantum properties certifiable in a device-independent manner remains, to our knowledge, an open problem.

In the {\em i.i.d.} setting, the PBR protocol is known to be asymptotically optimal (in terms of its confidence-gain rate). However, we see from \cref{fig:CHSH_lv3_ga099,fig:CGLMP_lv14_ga099_100k,fig:CHSH_extractability_ga099} that for a relatively small number of trials and with the {\em right choice} of the Bell function, the martingale-based protocol performs equally well, if not better. A similar observation was also noted in~\cite{WKCZ19} where the authors therein compare the PBR method with the Chernoff-Hoeffding bound in determining the success probability of Bernoulli trials. In our case, this is not surprising as the PBR method does not presuppose a Bell-like inequality but rather sacrifices some of the data to determine one. Indeed, if we equip the PBR protocol with the optimized Bell-like inequality right from the beginning, its performance is, as expected, no worse than the martingale-based protocol. See~\cref{fig:CHSH_lv3_ga099_fourmethods,fig:compareNvalCHSH100K,fig:compare Nval CGLMP,fig:CGLMP_lv14_ga099_100k_fourmethods,fig:CHSH_extractability_ga099_fourmethods,fig:compare Fval CHSH}
in~\cref{App:MISC} for some explicit examples.

Meanwhile, we also see from~\cref{fig:CHSH_confidence-gainrates-rel,fig:CGLMP02 compare gain rate L2,fig:CHSH_extractability_confidencegain} that for several cases that we have investigated, one's intuitive choice of the Bell function for the martingale-based method can lead to a relatively poor confidence-gain rate,  and hence impairs its efficiency to produce a good $p$-value bound, see~\cref{fig:compare Nval CGLMP,fig:compare Fval CHSH}. For example, even though the titled CHSH Bell inequality of~\cref{Ineq:TCHSH} is known to self-test {\em all} entangled two-qubit pure states $\ket{\psi(\theta)}$, this choice of the Bell function in the martingale-based method leads to a worse performance (for bounding the target-state-fidelity) compared with using the CHSH Bell function, which, in turn, gives a suboptimal performance compared with that derived from the PBR protocol, see~\cref{fig:CHSH_extractability_confidencegain}.
 At this point, it is worth reiterating that both protocols do {\em not} require the assumption that the experimental trials are {\em i.i.d}, even though we have only given, for simplicity, examples with {\em i.i.d.} trials.

Several research directions naturally follow from the present work. Firstly, there are the scalability questions: (1) how do the number of measurement bases and (2) the number of samples scale with the complexity (say, dimension) of the measured system? The former is again closely related to the general viability of the device-independent certification approach, where our understanding is far from complete. As for the latter, we remark it is indeed one of the goals of the present work to shed light on the sample complexity of our hypothesis-testing-based approaches. In some cases, such as the certification from the GHZ correlation, we see that hundreds of trials suffice, but in some others, several tens of thousands may be required to give a satisfactory level of certification. Still, some general understanding of how the sample size scales depend on the properties to be certified and the confidence level will be surely welcome.

Secondly, for experimental trials expected to deviate significantly from being {\em i.i.d.}, one should choose a much smaller block size $\Nint$ than the one adopted in our analysis using the PBR protocol. Intuitively, we should choose $\Nint$ so that the trials do not differ significantly within each block of data. In fact, for testing against LHV theories, some guidelines have been provided in~\cite{ZGK11} on choosing $\Nint$. A similar analysis for other DI certifications is clearly desirable. Next, even though our hypothesis-testing-based approaches enable rigorous DI certification with a confidence interval, by virtue of the techniques involved, one can only make a relatively {\em weak} certification: out of the many experimental trials, we can be sure that {\em at least one} consists of a setup that exhibits the desired property (say, with $99\%$ confidence). This is evidently far from satisfactory. A preferable certification scheme should allow one to comment on the general or {\em average} behavior of all the measured samples, as has been achieved in~\cite{BRS+21,GSD2022} for self-testing. 

Given that self-testing with high fidelity is technically challenging, it is still of interest to devise a {\em general recipe} for certifying the average behavior of other more specific properties (such as entanglement, steerability, etc.), which may already be sufficient for the specific information processing task at hand. However, note that the rejection of a null hypothesis on the {\em average behavior} (e.g., average negativity $\overline{\Neg(\rho)}\le \Neg_0$) necessarily entails the rejection of the corresponding null hypothesis  {\em for all trials} (e.g., $\Neg(\rho)\le \Neg_0$ in {\em every trial}). Thus, we may expect a tradeoff when switching from the current kind of hypothesis testing to that for an average behavior.

Also worth noting that if the {\em i.i.d.} assumption is somehow granted, then our protocols also certify the quality of the setup for every single run, including those that have not been measured. In this case, once a sufficiently small $p$-value bound is obtained, one can stop measuring the rest of the systems and use them, instead, for the information processing tasks of interest. Of course, since the {\em i.i.d.} assumption is generally not warranted, a protocol that achieves certification for some fraction of the copies while leaving the rest useful for subsequent tasks will be desirable. This has been considered for one-shot distillable entanglement in~\cite{AB19} and the self-testing fidelity in~\cite{GSD2022}. Again, a general treatment will be more than welcome (see, e.g.~\cite{Zhang:2023aa}).

\acknowledgments
We are grateful to Jean-Daniel Bancal, Gelo Tabia, Yanbao Zhang for many enlightening discussions, and to an anonymous referee for helpful suggestions. This work is supported by the National Science and Technology Council, Taiwan (Grants No. 107-2112-M-006-005-MY2 and 109-2112-M-006-010-MY3, 111-2112-M-005-007-MY4, 112-2119-M-001-004, 112-2119-M-001-006, 112-2124-M-002-003, 112-2628-M-006-007-MY4) and the Foxconn Research Institute, Taipei, Taiwan.
\end{CJK*}

\appendix
\section{Miscellaneous results}
\label{App:MISC}

In this Appendix, we provide some supplemental results to further illustrate the relative strength of the two hypothesis-testing protocols. For that matter, we extend some of the Figures shown in the main text to include two other plots in each of them. The first of this, dubbed ``PBR-simplified"  consists of the simplified implementation of the PBR protocol, where we use in \cref{Eq:KLD} and \cref{Eq:PBR} the relative frequency $\vecf$ computed from the raw data, instead of the regularized frequency $\vecf\reg$ . 

On the other hand, to see the best possible performance that one could hope for via any kind of implementation of the PBR protocol, we also replace $\vecf\reg$ in \cref{Eq:KLD} and \cref{Eq:PBR} by the ideal quantum correlation $\vecP_\Q$ used in the generation of the data. This gets rid of the statistical fluctuations involved in the estimation of $\vecP_\Q$ right from the beginning. We call this ``PBR-ideal" as it amounts to using the optimal Bell-like inequality right from the beginning.

\subsection{Negativity}
\label{Appdx.sec.CHSH}

\subsubsection{Certification based on the data generated from $\PCHSH$}

\begin{figure}[H]
    \centering
    \includegraphics[width=0.48\textwidth]{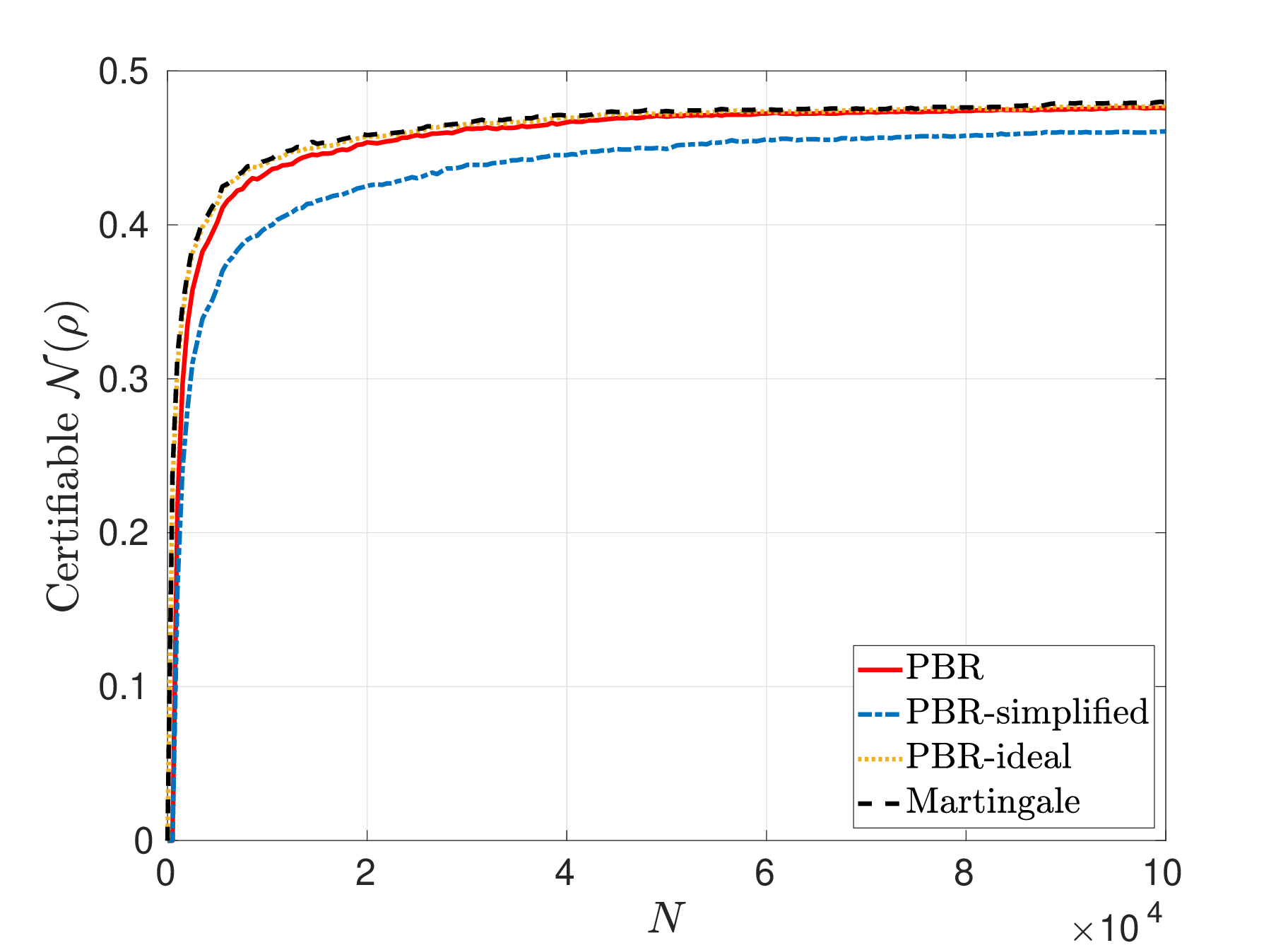}
    \caption{Extended plot from \cref{fig:CHSH_lv3_ga099} where we also include the results from the simplified implementation of the PBR protocol ``PBR-simplified" and the best that one could hope for in implementing the PBR protocol ``PBR-ideal".\label{fig:CHSH_lv3_ga099_fourmethods}
    }
\end{figure}

We start with \cref{fig:CHSH_lv3_ga099_fourmethods}, which extends~\cref{fig:CHSH_lv3_ga099}  by including the two plots mentioned above.

\begin{figure}[H]
    \begin{subfigure}{0.235\textwidth}
        \includegraphics[width=1\textwidth]{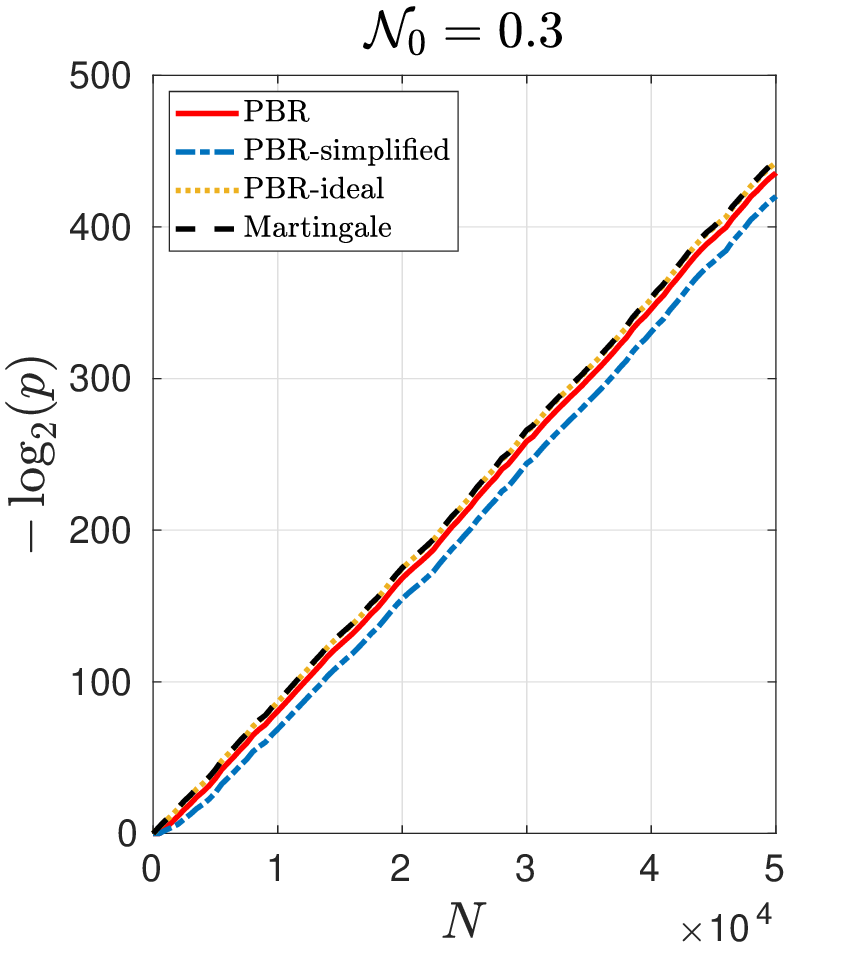}
        \caption{}
        \label{fig:minuslog2P_N03_lv3}
    \end{subfigure}
    \hfill
    \begin{subfigure}{0.235\textwidth}
        \includegraphics[width=1\textwidth]{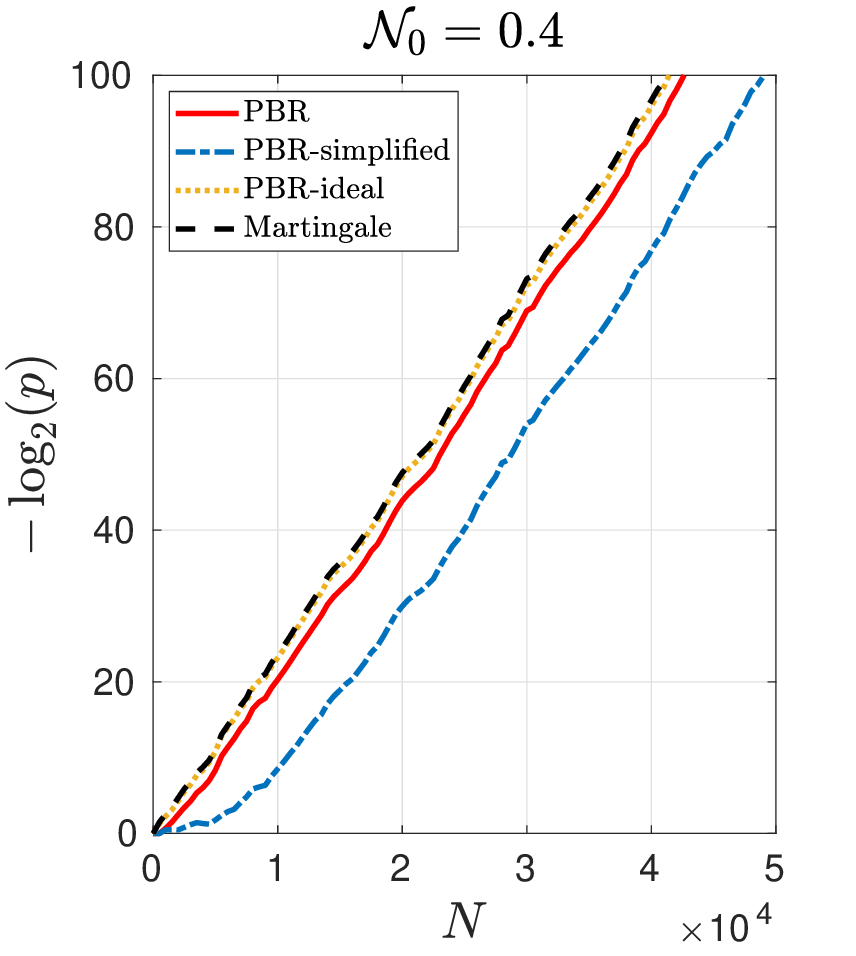}
        \caption{}
        \label{fig:minuslog2P_N04_lv3}
    \end{subfigure}
    \caption{Plot of -$\log2p\prot$ vs $N$ for Null Hypothesis~\ref{H:Negativity} with $\Neg_0=0.3$ and $0.4$. The parameters are the same as those described in~\cref{fig:CHSH_lv3_ga099}.  \label{fig:compareNvalCHSH100K}}
\end{figure}

Clearly, as we can see in \cref{fig:CHSH_lv3_ga099_fourmethods}, the performance of the PBR-simplified protocol is far worse than the PBR protocol. For example, after $N=10^5$ trials, the certifiable negativity is only about the same as that achieved with the PBR protocol with $N=2\times10^4$ trials. Similarly, we see in~\cref{fig:compareNvalCHSH100K} that with this protocol, the rate at which the $p$-value bound for Null Hypothesis~\ref{H:Negativity} with $\Neg_0=0.3$ and $0.4$ decreases at a rate that is far slower than the other protocols. On the other hand, we also see from these figures that the performance of PBR-ideal is essentially the same as the one given by the martingale-based protocol, an observation consistent with what one would expect from~\cref{fig:CHSH_confidence-gainrates-rel} (for $\theta=\frac{\pi}{4}$).

\subsubsection{Certification based on the data generated from $\vec{P}_\theta$}

\begin{figure}[H]
\centering
\includegraphics[width=0.48\textwidth]{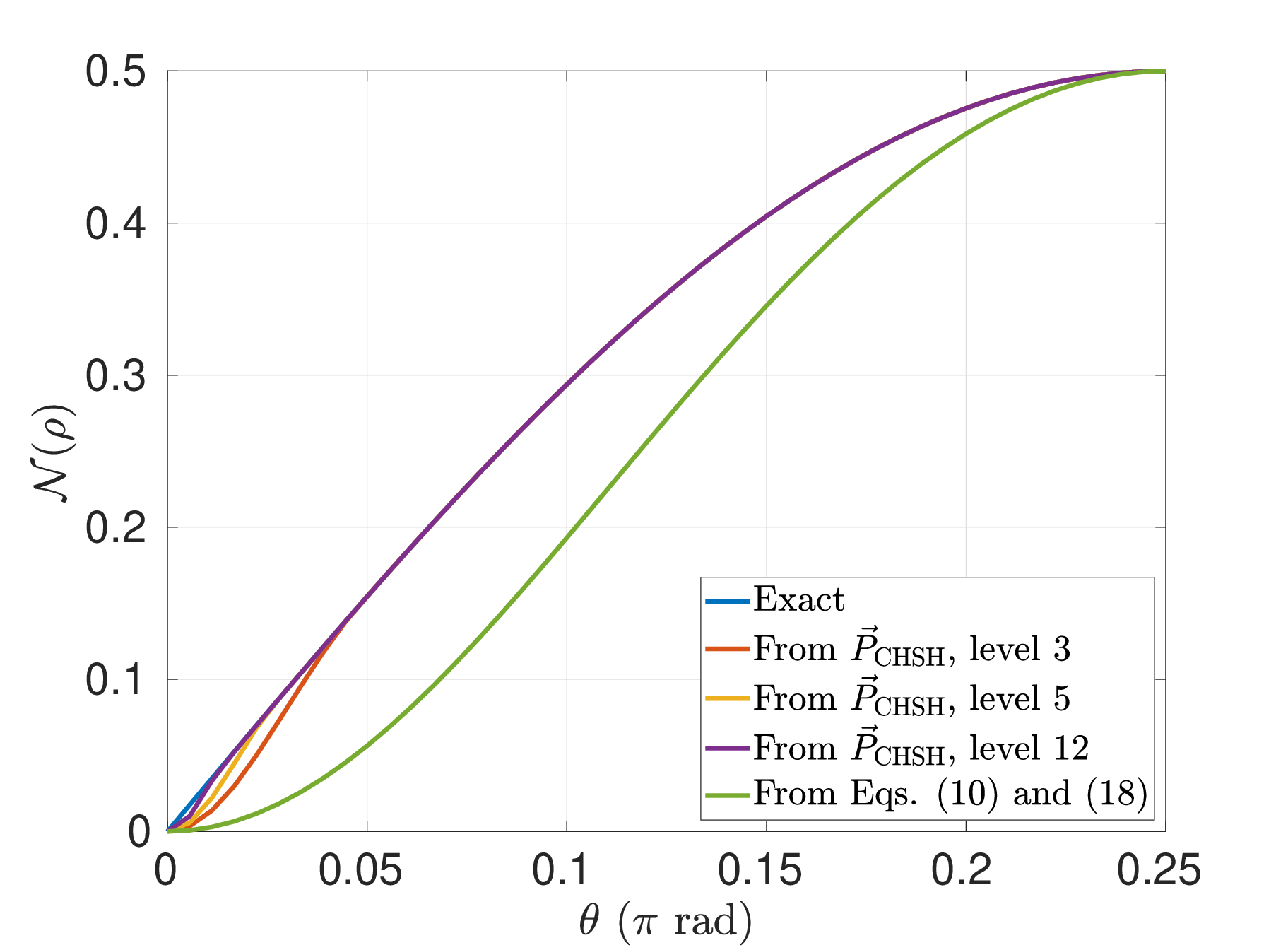} 
\caption{Comparison of the actual negativity (top) of $\psi(\theta)$, \cref{Eq:PsiTheta}, and their various device-independent lower bounds. 
The computation for the negativity lower bound based on the underlying correlation $\vecP_\theta$, derived from~\cref{Eq:Sefl-test:PES}, is obtained by solving \cref{Eq:SDP:MinNeg} whereas the bottom (dashed) line follows directly from the CHSH Bell inequality violation of these states, as given in \cref{Eq:CHSH_Neg,Eq:CHSH:General}.
}
\label{Fig:NegBound}
\end{figure}

In the main text, we mention, in relation to~\cref{fig:CHSH_confidence-gainrates-rel}, a possible cause of the noticeable discrepancy in the confidence-gain rates obtained between the two protocols. In~\cref{Fig:NegBound}, we see that, indeed, for any partially entangled two-qubit state $\ket{\psi(\theta)}$, the negativity lower bound obtained directly from the CHSH Bell violation, cf. ~\cref{Eq:CHSH_Neg}, is far from tight compared to that obtained by solving~\cref{Eq:SDP:MinNeg} for, say, level-3 of the hierarchy introduced in~\cite{MBL+13}. Interestingly, we also see from~\cref{Fig:NegBound} that negativity lower bound continues to improve even at level-12 of the hierarchy. See~\cite{LVL22} for closely related studies on the convergence of this and other SDP hierarchies.

\subsubsection{Certification based on the data generated from $\PCGLMP$}

\begin{figure}[H]
\centering
\includegraphics[width=0.48\textwidth]{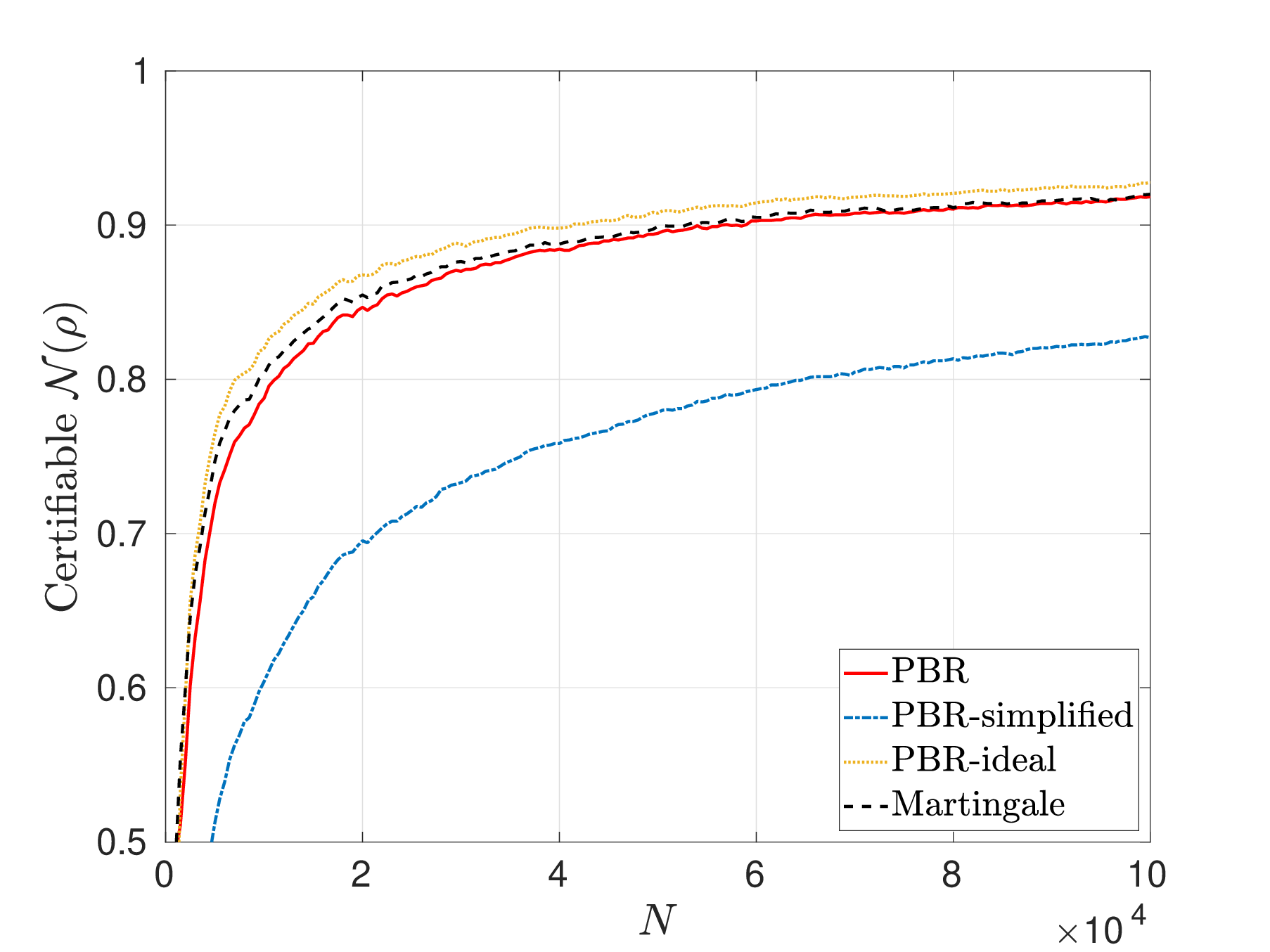} 
\caption{Extended plot from \cref{fig:CGLMP_lv14_ga099_100k} where we also include the results from the simplified implementation of the PBR protocol ``PBR-simplified" and the ideal protocol ``PBR-ideal".}
\label{fig:CGLMP_lv14_ga099_100k_fourmethods}
\end{figure}

The results analogous to~\cref{fig:CHSH_lv3_ga099_fourmethods} for the extension from ~\cref{fig:CGLMP_lv14_ga099_100k} can be found in  \cref{fig:CGLMP_lv14_ga099_100k_fourmethods}.

\begin{figure}[H]
    \begin{subfigure}{0.23\textwidth}
        \includegraphics[width=1\textwidth]{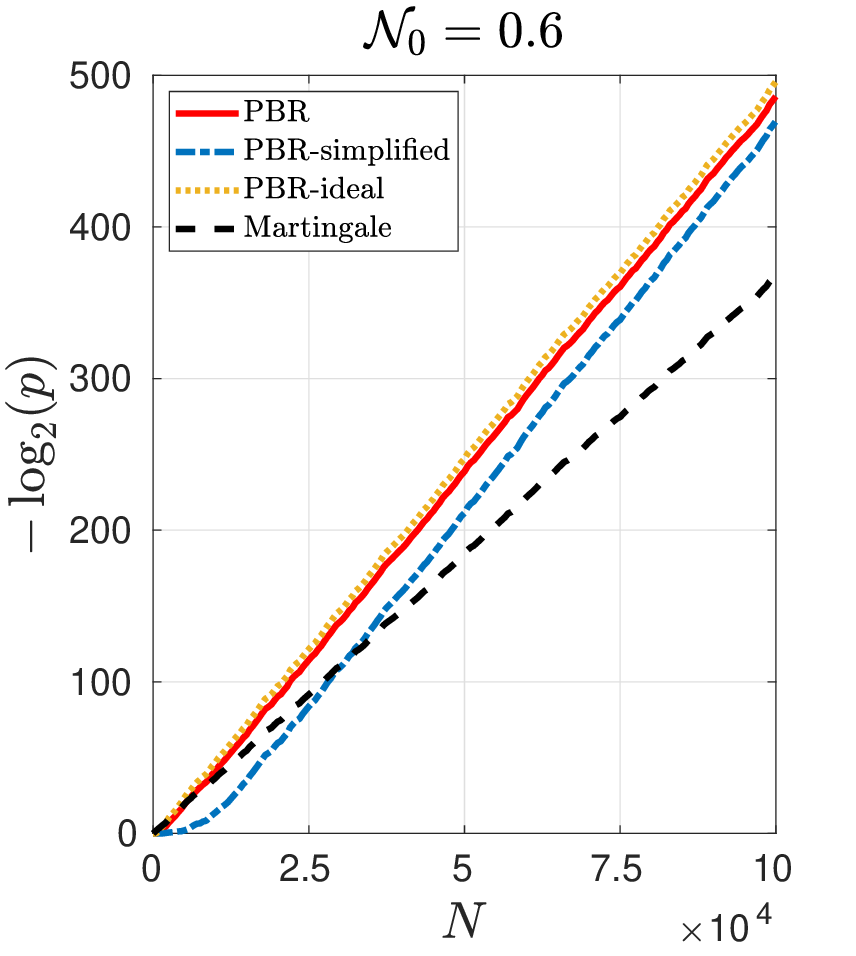}
        \caption{}
        \label{fig:minuslog2P_100k_N06_CGLMP}
    \end{subfigure}
    \hfill
    \begin{subfigure}{0.23\textwidth}
        \includegraphics[width=1\textwidth]{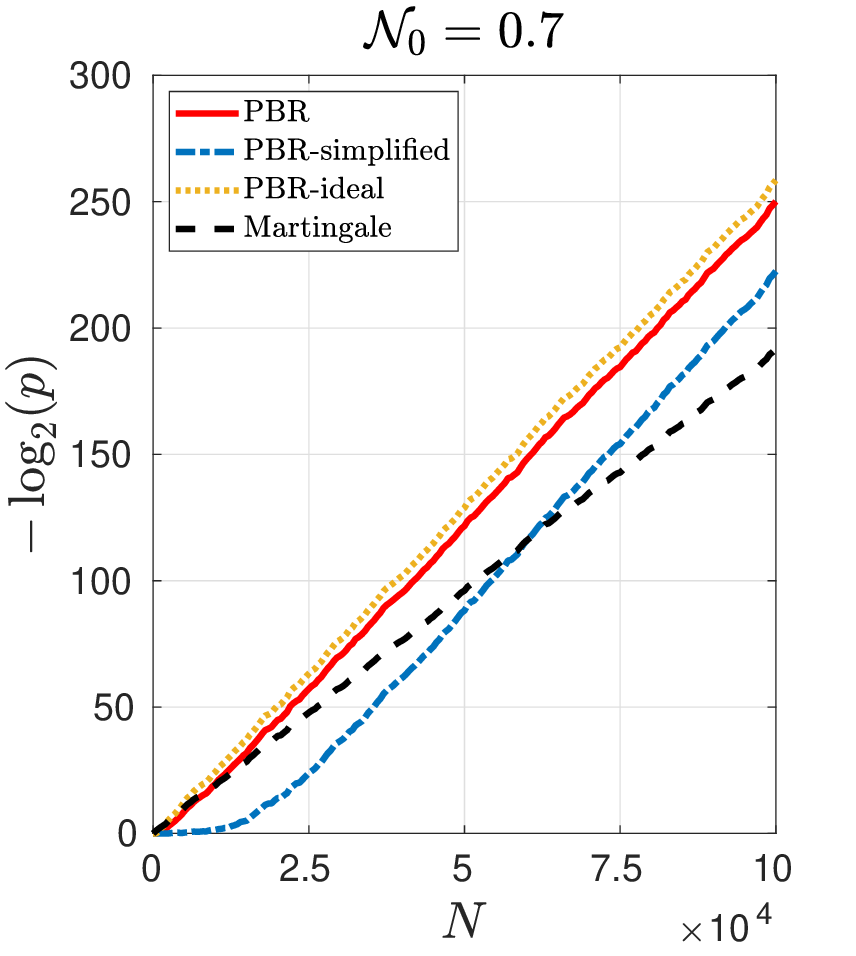}
        \caption{}
        \label{fig:minuslog2P_100k_N07_CGLMP}
    \end{subfigure}
    
    \medskip
    \begin{subfigure}{0.23\textwidth}
        \includegraphics[width=1\textwidth]{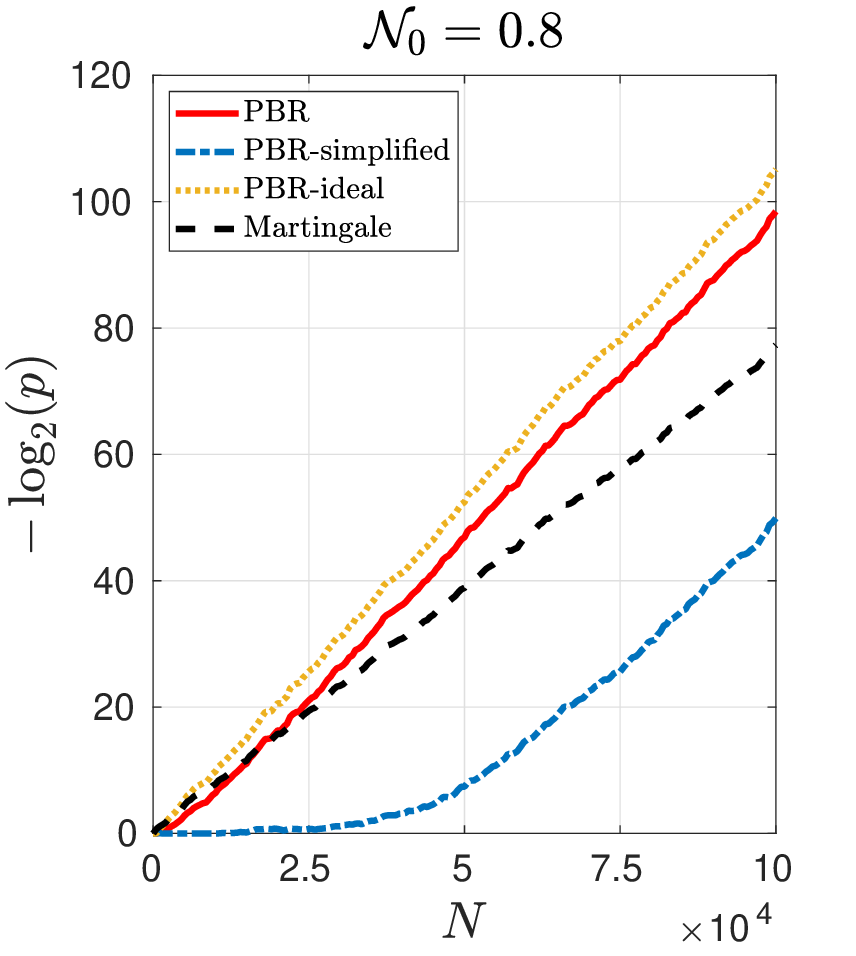}
        \caption{}
        \label{fig:minuslog2P_100k_N08_CGLMP}
    \end{subfigure}
    \hfill
    \begin{subfigure}{0.23\textwidth}
        \includegraphics[width=1\textwidth]{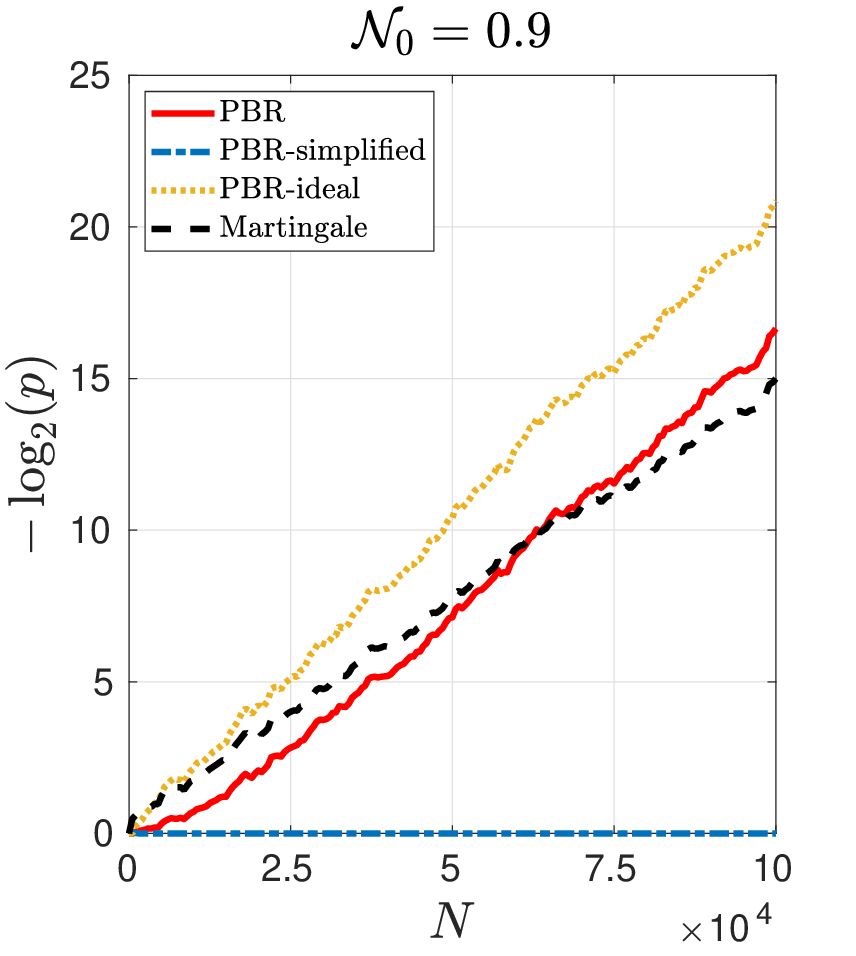}
        \caption{}
        \label{fig:minuslog2P_100k_N09_CGLMP}
    \end{subfigure}
    \caption{
    Plot of -$\log_2p\prot$ vs $N$ for hypothesis~\ref{H:Negativity} with $\Neg_0=0.6, 0.7, 0.8,$ and $0.9$. The parameters are the same as those described in~\cref{fig:CGLMP_lv14_ga099_100k}.  \label{fig:compare Nval CGLMP}}
\end{figure}
In this case, interestingly, we see that for $\Neg_0=0.6, 0.7$ and sufficiently large $N$, even the PBR-simplified protocol could overtake the martingale-based protocol in its $p$-value upper bound. As $\Neg_0$ increases, we also see that the difference in the performance between the PBR protocol and the PBR-ideal protocol becomes more pronounced. Still, for sufficiently large $N$, the PBR protocol eventually surpasses the martingale-based protocol in its $p$-value bound. This last observation is consistent with our observation in ~\cref{fig:CGLMP_lv14_ga099_100k_fourmethods} and the confidence gain rate shown in \cref{fig:CGLMP02 compare gain rate L2}.

\subsection{Fidelity}
\label{Appdx.sec.fidelity}

\begin{figure}[H]
\centering
\includegraphics[width=0.45\textwidth]{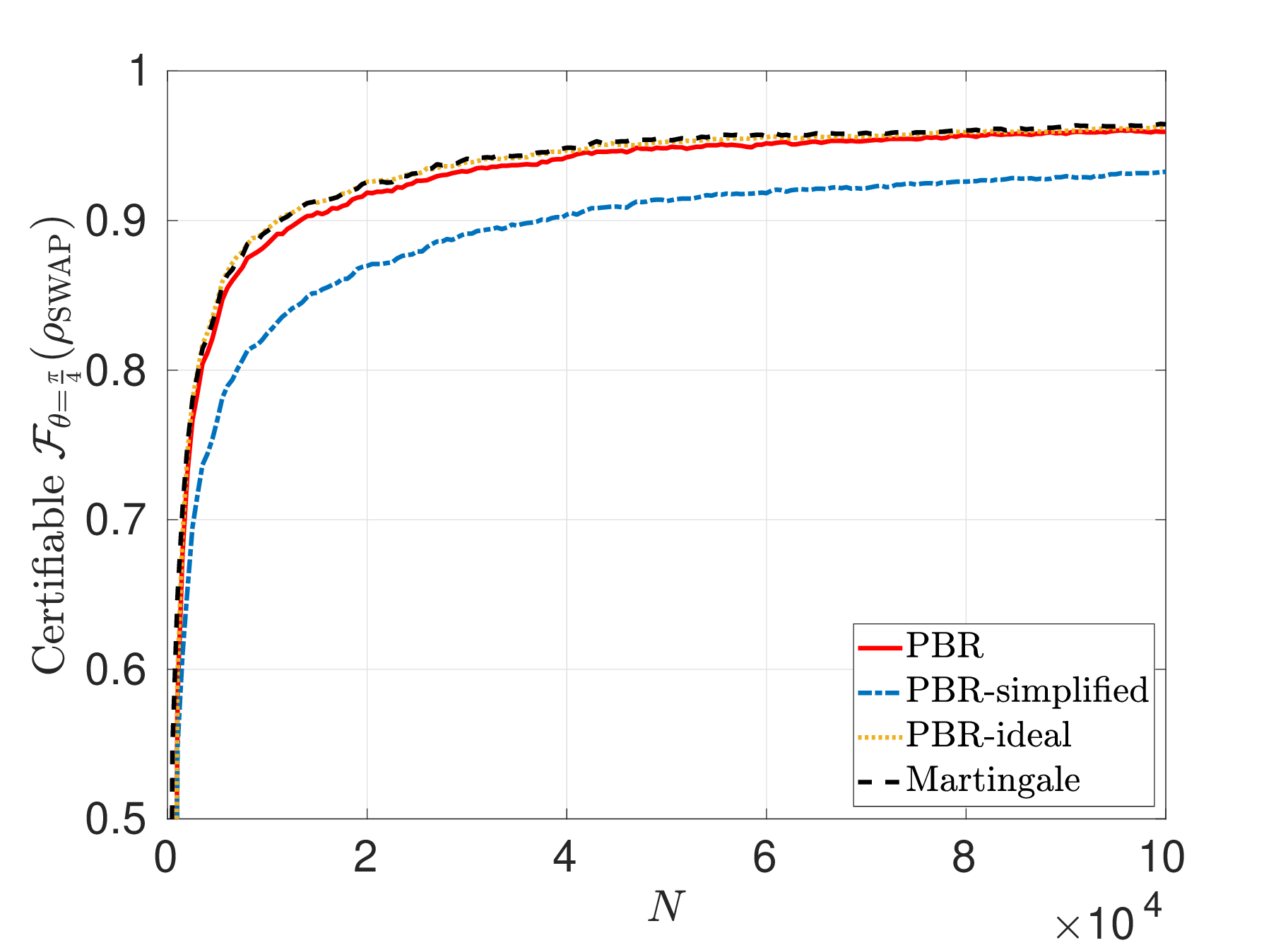} 
\caption{Extended plot from \cref{fig:CHSH_extractability_ga099} where we also include the results from the simplified implementation of the PBR protocol ``PBR-simplified" and the best that one could hope for in implementing the PBR protocol ``PBR-ideal".}
\label{fig:CHSH_extractability_ga099_fourmethods}
\end{figure}

\begin{figure}[H]
    \begin{subfigure}{0.23\textwidth}
        \includegraphics[width=1\textwidth]{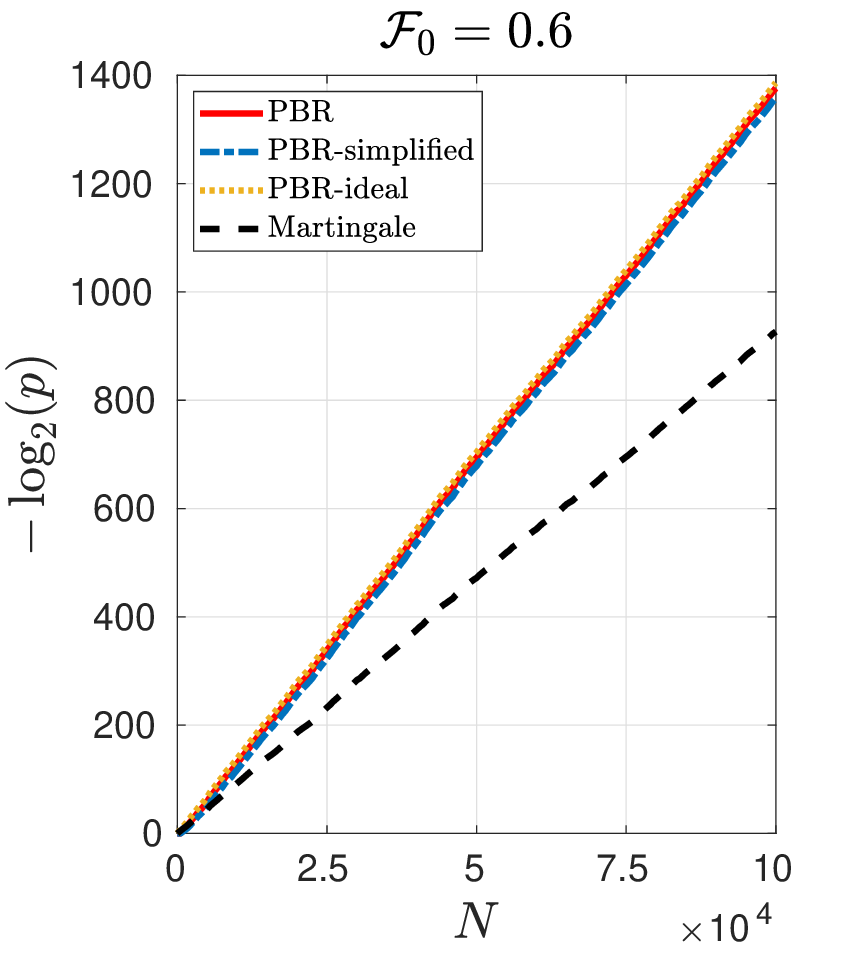}
        \caption{}
        \label{fig:minuslog2P_100k_F06_CHSH}
    \end{subfigure}
    \begin{subfigure}{0.23\textwidth}
        \includegraphics[width=1\textwidth]{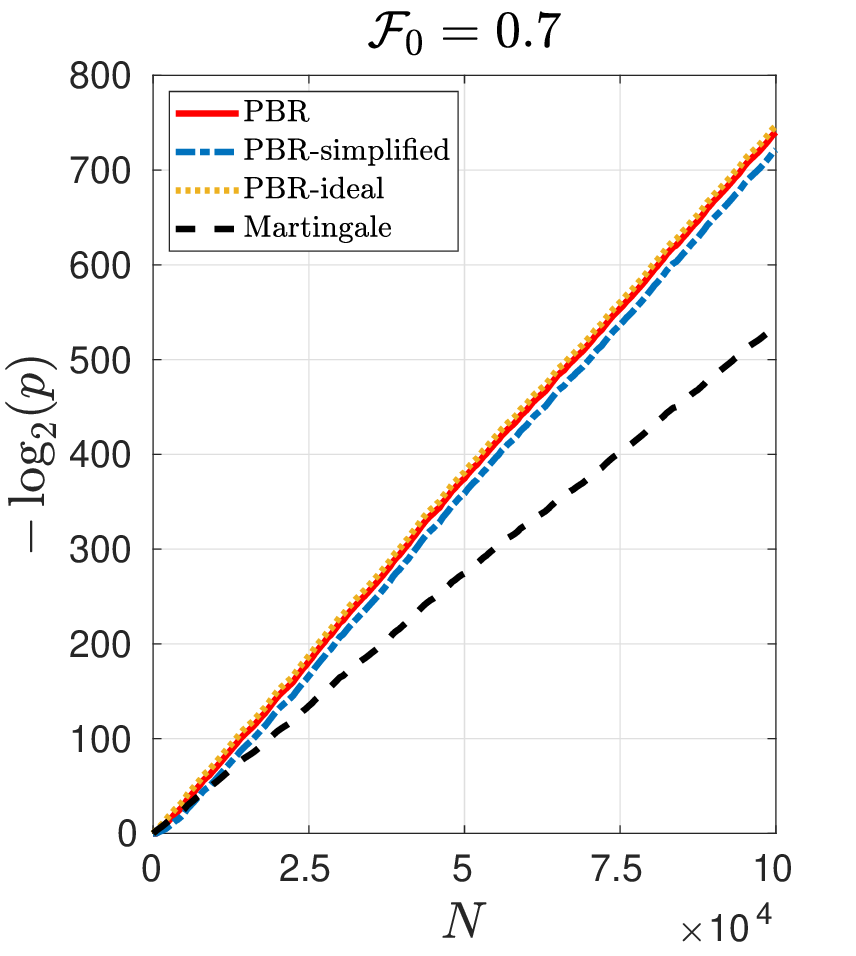}
        \caption{}
        \label{fig:minuslog2P_100k_F07_CHSH}
    \end{subfigure}
    \\
    \begin{subfigure}{0.23\textwidth}
        \includegraphics[width=1\textwidth]{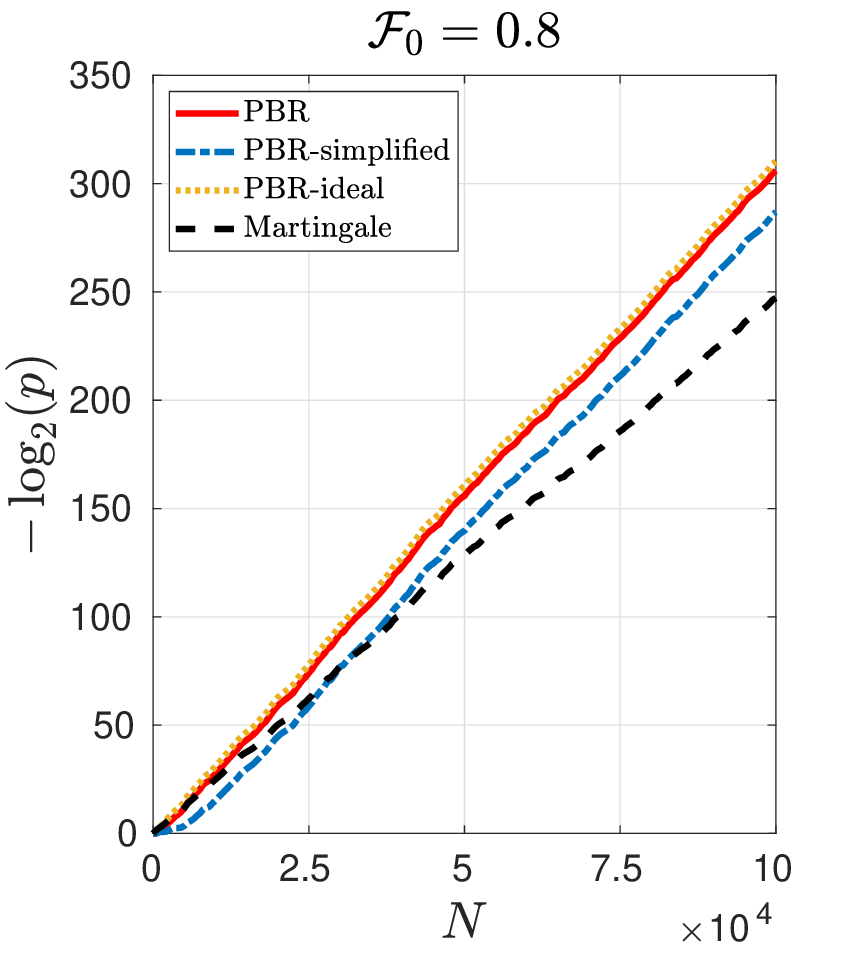}
        \caption{}
        \label{fig:minuslog2P_100k_F08_CHSH}
    \end{subfigure}
    \begin{subfigure}{0.23\textwidth}
        \includegraphics[width=1\textwidth]{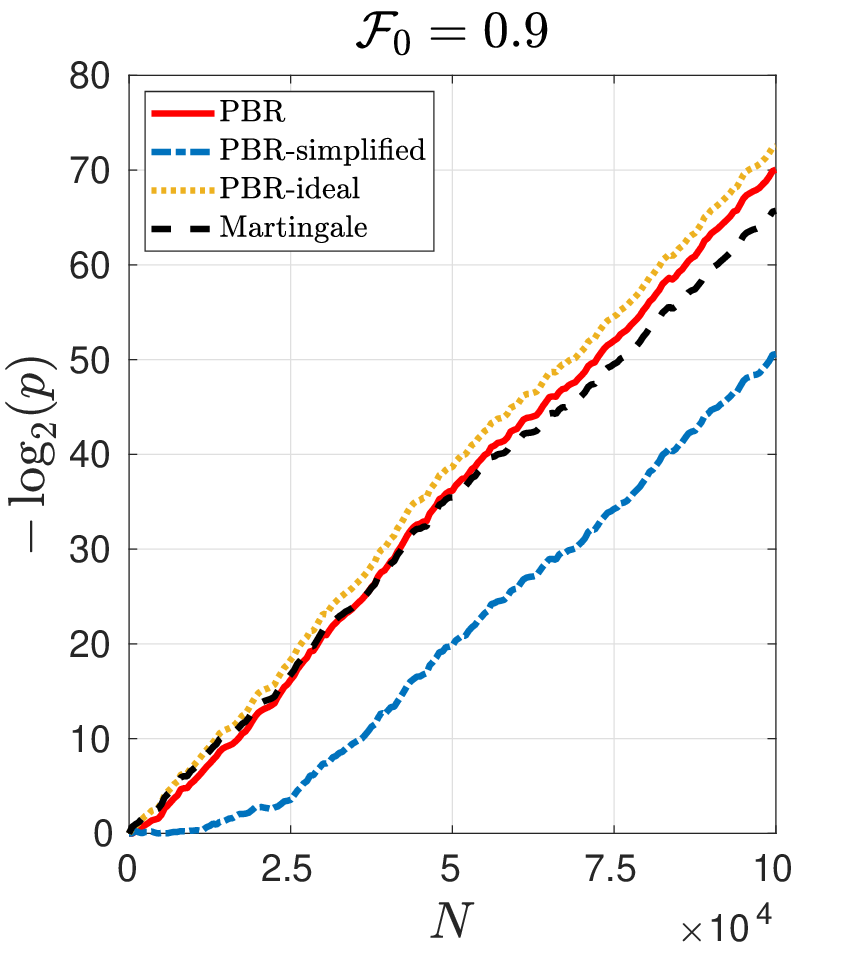}
        \caption{}
        \label{fig:minuslog2P_100k_F09_CHSH}
    \end{subfigure}
    \caption{
    Plot of -$\log_2p\prot$ vs $N$ for hypothesis~\ref{H:Fid} with $\F_0=0.6, 0.7, 0.8,$ and $0.9$. The parameters are the same as those described in~\cref{fig:CHSH_extractability_ga099}.  }    
    \label{fig:compare Fval CHSH}
\end{figure}

Next, let us include also the plots for PBR-ideal and PBR-simplified in~\cref{fig:CHSH_extractability_ga099}, as shown in~\cref{fig:CHSH_extractability_ga099_fourmethods}.
Again, we see that the performance of PBR is considerably impaired if we switch to PBR-simplified. Meanwhile, even though the difference between PBR-ideal and PBR is relatively insignificant for $\F_0=0.6,0.7$, and $0.8$ in \cref{fig:compare Fval CHSH}, we see that the difference is significant enough to be manifested in~\cref{fig:CHSH_extractability_ga099_fourmethods}.

\end{document}